\pdfoutput=1

\documentclass[11pt,twoside,a4paper,cmspaper,final,collab]{cms-tdr}
\def\svnVersion{199414}\def\cmsCernNo{2013-059}\def\cmsCernDate{\today}\def\cmsMessage{Published in the European Physical Journal C as \href{http://dx.doi.org/10.1140/epjc/s10052-013-2494-7}{\doi{10.1140/epjc/s10052-013-2494-7.}}}

\begin{document}\cmsNoteHeader{TOP-11-027}

\hyphenation{had-ron-i-za-tion}
\hyphenation{cal-or-i-me-ter}
\hyphenation{de-vices}
\RCS$HeadURL: svn+ssh://svn.cern.ch/reps/tdr2/papers/TOP-11-027/trunk/TOP-11-027.tex $
\RCS$Id: TOP-11-027.tex 191002 2013-06-18 00:30:27Z alverson $
\newlength\cmsFigWidth
\ifthenelse{\boolean{cms@external}}{\setlength\cmsFigWidth{0.95\columnwidth}}{\setlength\cmsFigWidth{0.48\textwidth}}
\ifthenelse{\boolean{cms@external}}{\providecommand{\cmsLeft}{top\xspace}}{\providecommand{\cmsLeft}{left\xspace}}
\ifthenelse{\boolean{cms@external}}{\providecommand{\cmsRight}{bottom\xspace}}{\providecommand{\cmsRight}{right\xspace}}
\renewcommand{\vec}[1]{\mathbf{#1}}
\newcommand{\vPTm}{\vec\PTm}
\renewcommand{\Pl}{\ensuremath{\ell}}
\newcommand{\ourlumi}{5.0}
\newcommand{\oursyst}{^{+1.7}_{-2.1}}
\providecommand{\mct}{\ensuremath{M_{\mathrm{CT}}}}
\providecommand{\mtt}{\ensuremath{M_{\mathrm{T}2}}}
\providecommand{\mctp}{\ensuremath{M_{\mathrm{CT}\perp}}}
\providecommand{\mttp}{\ensuremath{M_{\mathrm{T}2\perp}}}
\providecommand{\mttpaOLD}{\ensuremath{M_{\mathrm{T}2\perp}^{210}}}
\providecommand{\mttpbOLD}{\ensuremath{M_{\mathrm{T}2\perp}^{221}}}
\providecommand{\mttpamaxOLD}{\ensuremath{M^\text{max}_{\mathrm{T}2\perp}({210})}}
\providecommand{\mttpbmaxOLD}{\ensuremath{M^\text{max}_{\mathrm{T}2\perp}({221})}}
\providecommand{\ellell}{\ensuremath{ {\ell\ell}  }}
\providecommand{\beebee}{\ensuremath{ {\cPqb\cPqb}}}
\providecommand{\bell}  {\ensuremath{ {\cPqb\Pl}  }}
\providecommand{\mttpa}{\ensuremath{\mu_{\ellell}}}
\providecommand{\mttpb}{\ensuremath{\mu_{\beebee}}}
\providecommand{\mttpamax}{\ensuremath{\mu^{\max}_{\ellell}}}
\providecommand{\mttpbmax}{\ensuremath{\mu^{\max}_{\beebee}}}
\providecommand{\mttpc}{\ensuremath{M_{\mathrm{T}2\perp}^{220}}}
\providecommand{\mctpa}{\ensuremath{M_{\mathrm{CT}\perp}^{210}}}
\providecommand{\mctpb}{\ensuremath{M_{\mathrm{CT}\perp}^{221}}}
\providecommand{\mctpc}{\ensuremath{M_{\mathrm{CT}\perp}^{220}}}
\providecommand{\mbl}{\ensuremath{M_{\cPqb\Pl}}}
\providecommand{\MT}{\ensuremath{M_{\mathrm{T}}}}
\providecommand{\MC}{\ensuremath{M_C}}
\providecommand{\testmass}{\ensuremath{\widetilde{M}_C}}
\providecommand{\testmassnu}{\ensuremath{\widetilde{m}_\cPgn}}
\providecommand{\testmassW}{\ensuremath{\widetilde{M}_\PW}}
\providecommand{\mctmax}{\ensuremath{M^\text{max}_{\mathrm{CT}}}}
\providecommand{\mttmax}{\ensuremath{M^\text{max}_{\mathrm{T}2}}}
\providecommand{\mctpmax}{\ensuremath{M^\text{max}_{\mathrm{CT}\perp}}}
\providecommand{\mttpmax}{\ensuremath{M^\text{max}_{\mathrm{T}2\perp}}}
\providecommand{\mttpcmax}{\ensuremath{M^\text{max}_{\mathrm{T}2\perp}({220})}}
\providecommand{\mblmax}{\ensuremath{M^\text{max}_{\cPqb\Pl}}}
\providecommand{\Ea}{\ensuremath{E_{210}}}   
\providecommand{\Eb}{\ensuremath{E_{221}}}   
\providecommand{\Ec}{\ensuremath{E_{M\cPqb\Pl}}}
\providecommand{\mt}{\ensuremath{M_{\cPqt}}}
\providecommand{\mw}{\ensuremath{M_{\PW}}}
\providecommand{\mn}{\ensuremath{m_{\cPgn}}}
\providecommand{\mb}{\ensuremath{m_{\cPqb}}}
\providecommand{\mtsq}{\ensuremath{M^2_{\cPqt}}}
\providecommand{\mwsq}{\ensuremath{M^2_{\PW}}}
\providecommand{\mnsq}{\ensuremath{m^2_{\cPgn}}}
\providecommand{\mbsq}{\ensuremath{m^2_{\cPqb}}}
\providecommand{\Pt}{\ensuremath{P_{\text{T}}}}
\providecommand{\pt}{\ensuremath{p_{\text{T}}}}
\providecommand{\vPt}{\ensuremath{\vec{P}_{\text{T}}}}
\providecommand{\vpt}{\ensuremath{\vec{p}_{\text{T}}}}
\providecommand{\Ptp}{\ensuremath{P_{\text{T}\perp}}}
\providecommand{\ptp}{\ensuremath{p_{\text{T}\perp}}}
\providecommand{\vPtp}{\ensuremath{\vec{P}_{\text{T}\perp}}}
\providecommand{\vptp}{\ensuremath{\vec{p}_{\text{T}\perp}}}

\newcommand{\nuPhi}{\ensuremath{{\Phi}}}  
\newcommand{\oPhi}{\ensuremath{\phi_{\text{US}}}}   
\newcommand{\myhalf}{\ensuremath{\tfrac{1}{2}}}
\newcommand{\delphi}{\ensuremath{\tfrac{1}{2}(\phi_1-\phi_2)}}
\newcommand{\Dphi}{\ensuremath{\Delta\phi}}
\newcommand{\hDphi}{\ensuremath{\tfrac{1}{2}\Delta\phi}}
\newcommand{\hpi}{\ensuremath{\tfrac{\pi}{2}}}
\newcommand{\ptpt}{\ensuremath{p_{\text{T1}}p_{\text{T2}}}}
\providecommand{\vpta}{\ensuremath{\vec{p}_{\text{T1}}}}
\providecommand{\vptb}{\ensuremath{\vec{p}_{\text{T2}}}}

\numberwithin{equation}{section}
\numberwithin{figure}{section}

\cmsNoteHeader{AN-10-316} 

\title{Measurement of masses in the \ttbar system by kinematic endpoints in pp collisions at $\sqrt{s}=7\TeV$}

\date{\today}

\abstract{
A simultaneous measurement of the top-quark, \PW-boson, and neutrino masses
is reported for $\ttbar$ events selected in the dilepton
final state from a data sample corresponding to an integrated luminosity of
5.0\fbinv collected by the CMS experiment in pp collisions at
$\sqrt{s}=7$\TeV. The analysis is based on endpoint determinations in
kinematic distributions. When the neutrino and W-boson masses are
constrained to their world-average values, a top-quark mass value of
$M_{\mathrm{t}}=173.9\pm 0.9\stat{}\oursyst\syst\GeV$ is
obtained. When such constraints are not used, the three particle masses are
obtained in a simultaneous fit. In this unconstrained mode the study serves
as a test of mass determination methods that may be used in beyond standard
model physics scenarios where several masses in a decay chain may be
unknown and undetected particles lead to underconstrained kinematics.}

\hypersetup{%
pdfauthor={CMS Collaboration},%
pdftitle={Measurement of masses in the t t-bar system by kinematic endpoints in pp collisions at sqrt(s)=7 TeV},%
pdfsubject={CMS},%
pdfkeywords={CMS, physics, top quark}}
\maketitle 

\section{Introduction}\label{s:intro}

The determination of the top-quark mass sets a fundamental benchmark for
the standard model (SM), and is one of the precision measurements that
defines electroweak constraints on possible new physics beyond the
SM~\cite{Flacher:2008zq}.  With the recent
observations~\cite{:2012gk,:2012gu} of a Higgs boson candidate at a mass of
approximately 125\GeV, existing data can now overconstrain the SM. The top
quark plays an important role in such constraints because its large mass,
appearing quadratically in loop corrections to many SM observables,
dominates other contributions. It is also key to the quartic term in the
Higgs potential at high energy, and therefore to the question of stability
of the electroweak vacuum~\cite{Degrassi:2012ry,Alekhin:2012py}. For these
reasons, precise top-quark mass determinations are essential to
characterize and probe the SM. Recent results obtained at the Large Hadron
Collider (LHC) for the top-quark mass in \ttbar events include those
reported by ATLAS \cite{ATLAS:2012aj},
$\mt=174.5 \pm 0.6\stat \pm 2.3\syst\GeV$,
and  by the Compact Muon Solenoid (CMS) \cite{CMSlepjets},
$\mt=173.49\pm 0.43\stat \pm 0.98\syst\GeV$,
using the semileptonic decay channel of the \ttbar\ pair. The CMS
Collaboration has also reported a measurement \cite{CMSdilepton} in the
dilepton channel,
$\mt=172.5 \pm 0.4\stat \pm 1.5\syst\GeV$.
A recent summary of top-quark mass measurements conducted by the CDF and D0
Collaborations~\cite{Aaltonen:2012ra} reports a combined result
$\mt=173.18 \pm 0.56\stat\pm 0.75\syst\GeV$.

In parallel with recent measurements of the properties of the top quark at
the LHC, there has been a great deal of theoretical progress on methods
using endpoints of kinematic variables to measure particle masses with
minimal input from simulation. These methods are generally aimed at
measuring the masses of new particles, should they be discovered, but can
also be applied to measure the masses of standard model particles such as
the top quark. Such an application acts as both a test of the methods and a
measurement of the top-quark mass utilizing technique very different from
those used in previous studies.

Indeed, top-quark pair production provides a good match to these new
methods, as dilepton decays of top-quark pairs ($\cPqt\cPaqt\to
(\cPqb\Pl^+\nu)(\cPaqb \Pl^-\bar\nu)$) provide challenges in mass
measurement very similar to the ones that these methods were designed to
solve. A key feature of many current theories of physics beyond the
standard model is the existence of a candidate for dark matter, such as a
weakly interacting massive particle (WIMP). These particles are usually
stabilized in a theory by a conserved parity, often introduced {\em ad
hoc}, under which SM particles are even and new-physics particles are odd.
Examples include $R$-parity in supersymmetry (SUSY) and $T$-parity in
little-Higgs models. One consequence of this parity is that new physics
particles must be produced in pairs. Each of the pair-produced particles
will then decay to a cascade of SM particles, terminating with the lightest
odd-parity particle of the new theory. In such cases, there will be two
particles which do not interact with the detector, yielding events where
the observable kinematics are underconstrained. Mass measurements in these
events are further complicated by the presence of multiple new particles
with unknown masses.

\begin{figure*}[htb]
    \begin{center}
    \includegraphics[width=0.9\textwidth]{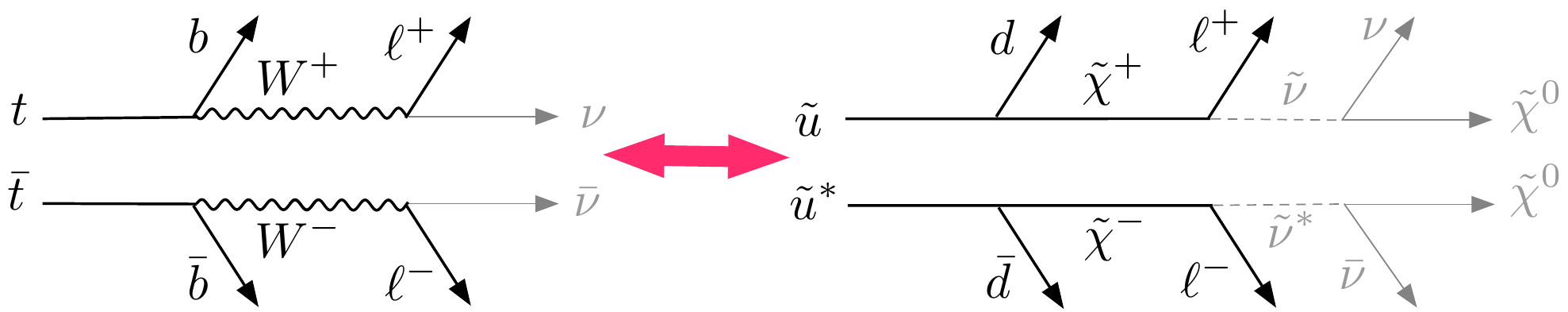}
    \caption{
    Top-quark pair dilepton decays, with two jets, two leptons, and two
    unobserved particles (left) exhibit a signature similar to some SUSY
    modes (right). In the figure, $\widetilde{u}$, $\PSGc^\pm$, $\widetilde{\nu}$, and $\PSGcz$ denote the u-squark, chargino, sneutrino, and
    neutralino respectively; an asterisk indicates the antiparticle of the
    corresponding SUSY particle.
    }
    \label{f:topsusy}
    \end{center}
\end{figure*}

The dilepton decays of \ttbar\ events at the LHC offer a rich source of
symmetric decay chains terminating in two neutrinos.  With their
combination of jets, leptons, and undetected particles, these \ttbar events
bear close kinematic and topological resemblance to new-physics scenarios
such as the supersymmetric decay chain illustrated in Fig.~\ref{f:topsusy}.
 This correspondence has motivated~\cite{Burns:2008va} the idea to use the
abundant \ttbar samples of the LHC as a testbed for the new methods and
novel observables that have been proposed to handle mass measurement in
new-physics events~\cite{barr2010review}. A simultaneous measurement of the
top-quark, W-boson, and neutrino masses in dilepton \ttbar decays closely
mimics the strategies needed for studies of new physics.

The analysis presented here focuses on the \mtt\ variable and its
variants~\cite{lester1999measuring,barr2010review}. These kinematic
observables are mass estimators that will be defined in Sec.
\ref{sec:method}. The goals of this analysis are two-fold: to demonstrate
the performance of a new mass measurement technique, and to make a precise
measurement of the top-quark mass. To demonstrate the performance of the
method, we apply it to the \ttbar\ system assuming no knowledge of the
W-boson or neutrino masses. This allows us to measure the masses of all
three undetected particles involved in the dilepton decay: the top quark, W
boson, and neutrino. This ``unconstrained'' fit provides a test of the
method under conditions similar to what one might expect to find when
attempting to measure the masses of new particles. In order to make a
precise measurement of the top-quark mass, on the other hand, we assume the
world-average values for the W boson and neutrino masses. This
``doubly-constrained'' fit achieves a precision in the top-quark mass
determination similar to that obtained by traditional methods.
The \mtt\ observable has been previously suggested~\cite{Cho:2008cu}
or used~\cite{Aaltonen:2009rm}
in top-quark mass measurements.

In considering any top-quark mass measurement, however, it is critical to confront
the fact that deep theoretical problems complicate the interpretation of
the measurement. The issues arise because a top quark is a colored object
while the W boson and hadronic jet observed in the final state are not. In
the transition $\cPqt\to\PW\cPqb$, a single color charge must come from
elsewhere to neutralize the final-state b jet, with the inevitable
consequence that the observed energy and momentum of the final state differ
from that of the original top quark. The resulting difference between
measured mass and top-quark mass is therefore at least at the level at
which soft color exchanges occur, \ie
${\sim}\Lambda_{\text{QCD}}$~\cite{Bigi:1994em,Smith}. In
the current state of the art, a Monte Carlo (MC) generator is normally used
to fix a relationship between the experimentally measured mass of the final
state and a top-quark mass parameter of the simulation; but model
assumptions upon which the simulation of nonperturbative physics depend
further limit the precision of such interpretative statements to about
1\GeV~\cite{Hoang:2008xm}.

We therefore take care in this measurement to distinguish between the
interpretive use of MC simulation described above, which is inherently
model dependent, and experimental procedures, which can be made clear and
model independent. A distinctive feature of the top-quark mass measurement
reported here is its limited dependence on MC simulation. There is no
reliance on MC templates~\cite{Aaltonen:2009rm}, and the endpoint method gives a result which is
consistent with the kinematic mass in MC without further tuning or
correction. For this reason, the measurement outlined here complements the
set of conventional top-quark mass measurements, and is applicable to
new-physics scenarios where MC simulation is used sparingly.

\section{The CMS Detector and Event Reconstruction} 
\label{sec:the_cms_detector}
The central feature of the CMS apparatus is a superconducting solenoid of
6\unit{m} internal diameter, providing a magnetic field of 3.8\unit{T}.
Inside the superconducting solenoid volume are silicon pixel and strip
trackers, a lead tungstate crystal electromagnetic calorimeter, and a
brass/scintillator hadron calorimeter. Muons are measured in gas-ionization
detectors embedded in the steel flux return yoke. Extensive forward
calorimetry complements the coverage provided by the barrel and endcap
detectors. A more detailed description of the CMS detector can be found in
Ref.~\cite{Chatrchyan:2008zzk}.

Jets, electrons, muons, and missing transverse momentum are reconstructed
using a global event reconstruction technique, also called particle-flow
event reconstruction~\cite{CMS-PAS-PFT-09-001,CMS-PAS-PFT-10-001}. Hadronic
jets are clustered from the reconstructed particles with the infrared and
collinear-safe anti-\kt algorithm~\cite{Cacciari:2008gp}, using
a size parameter 0.5. The jet momentum is determined as the vectorial sum
of all particle momenta in this jet, and is found in the simulation to be
within 5\% to 10\% of the true momentum over the whole transverse momentum
(\pt) spectrum and detector acceptance. Jet energy corrections are derived
from the simulation, and are confirmed in measurements on data with the
energy balance of dijet and photon+jet events~\cite{Chatrchyan:2011ds}. The
jet energy resolution amounts typically to 15\% at jet \pt\ of 10\GeV, 8\%
at 100\GeV, and 4\% at 1\TeV. The missing transverse momentum vector is
defined by $\vPTm\equiv-\sum\vpt$ where the sum is taken over all
particle-flow objects in the event; and missing transverse ``energy'' is
given by $\MET\equiv|\vPTm|$.

\section{Event Selection}

The data set used for this analysis corresponds to an integrated luminosity
of 5.0\fbinv of proton-proton collisions at $\sqrt{s}=7\TeV$ recorded by
the CMS detector in 2011. We apply an event selection to isolate a dilepton
sample that is largely free of  backgrounds. We require two well-identified
and isolated opposite-sign leptons (electrons or muons) passing dilepton
trigger requirements; the minimum \pt\ requirements for the triggers are
17\GeV and 8\GeV for the leading and sub-leading leptons. In addition we
require at least two b-tagged jets, subsequently used in the top-quark
reconstruction, and missing transverse energy. Here and throughout this
paper, we use $\Pl$ (and ``lepton'') to denote an electron or muon; the
signal decays of interest are $\cPqt\to \cPqb\Pl\cPgn$. Leptons must
satisfy $\pt>20\GeV$ and the event is vetoed if the leptons have the same
flavor and their dilepton invariant mass is within 15\GeV of the Z boson
mass. If three leptons are found, the two highest-\pt\ leptons forming an
opposite-sign pair are selected. Jets must satisfy $\pt>30\GeV$ after
correcting for additive effects of pileup (multiple proton collisions in a
single crossing) and multiplicative effects of jet energy scale
calibration. Jets are further required to lie within $|\eta|<2.5$, where
$\eta$ is the pseudorapidity variable, $\eta\equiv -\ln[\tan(\theta/2)]$.
The b-tagging algorithm is the Combined Secondary Vertex (CSV) tagger of
Ref. \cite{BTV-11-004}, deployed here with an operating point that yields a
tagging efficiency of $85\%$ and mistag rate of $10\%$.
The mistag rate measures the probability for a light quark or gluon jet
to be misidentified as a b jet.
In the subsample of events passing all selection requirements of
this analysis the b-jet purity is $91\%$.
Jet masses are required to satisfy a very loose requirement
$m_\text{jet}< 40\GeV$  to assure the existence of kinematic solutions and
reject poorly reconstructed jets. The missing transverse energy must
satisfy $\MET > 30$\GeV for $\Pep\Pem$ and $\Pgmp\Pgmm$ events and
$\MET>20$\GeV for $\Pe^\pm\Pgm^\mp$ events, where Drell--Yan backgrounds
are smaller. With the exception of the b-tagging criteria and the b-jet
mass requirement, all selection requirements summarized here are discussed
in more detail in \cite{Chatrchyan:2012bra,Chatrchyan:2012ea}. The sample
of events in data meeting all of the signal selection criteria contains
8700 events.

\section{Kinematic Variables}
\label{sec:method}

The endpoint method of mass extraction is based on several variables that
are designed for use in the kinematically complex environment of events
with two cascade decays, each ending in an invisible particle. The
challenge here is two-fold, combining the complications of a many-body
decay with the limitations of an underconstrained system. In a two-body
decay $A\to~B~C$, the momentum of either daughter in the parent rest frame
exhibits a simple and direct relationship to the parent mass. In a
three-body decay, $A\to~B~C~D$, the relationship is less direct, encoded
not in a delta function of momentum but in the kinematic boundary of the
daughters' phase space.  In general, the parent mass may be determined from
the endpoints of the observable daughter momenta in the parent rest frame.
To carry out this program, however, the daughter masses must be known and
enough of the momenta be measurable or constrained by conservation laws to
solve the kinematic equations.

Applying this program to the measurement of the top-quark mass in the decay
$\cPqt\to \cPqb\Pl\cPgn$, one immediately encounters a number of obstacles.
At a hadron collider, the \ttbar\ system is produced with unknown
center-of-mass energy and has an event-dependent \PT-boost due to recoil
from the initial-state radiation (ISR). Furthermore, in pp collisions we
can apply constraints of momentum conservation only in the two dimensions
transverse to the beam direction. Since top quarks are normally produced in
pairs, the individual neutrino momenta are indeterminate, adding further
complication. These obstacles seem daunting but can be overcome by the use
of ``designer'' kinematic variables  \mtt\ \cite{lester1999measuring} and
\mct\ \cite{Tovey:2008ui}, which,  by construction, address precisely these
issues. In this paper we use \mtt. Because the transverse momentum of the
\ttbar system varies from event to event, the \PT-insensitive version
\cite{Konar:2009wn,Matchev:2009ad}, \mttp, is particularly useful.  To
measure the masses of the top-quark, W-boson, and neutrino, we measure the
endpoints of three kinematic distributions, \mttpa, \mttpb, and \mbl, as
discussed in the following  subsections.

\subsection{\texorpdfstring{\mtt}{} and Subsystem Variables}

\subsubsection{The \texorpdfstring{\mtt}{} Observable}
\label{sec:mttobservable}
The variable \mtt\ is based on the transverse mass, \MT, which was first
introduced to measure the W-boson mass in the decay $\PW\to \Pl\nu$.  In
this case, \MT\ is defined by
\begin{equation}
	\MT^{2} \equiv \mn^{2} + m_{\Pl}^{2} +
	2(\ET^{\cPgn}\ET^{\Pl} -
	\vpt^{\cPgn}\cdot \vpt^{\Pl}).
	\label{e:mt}
\end{equation}
The observable \MT\ represents the smallest mass the \PW\ boson could have
and still give rise to the observed transverse momenta $\vpt^{\Pl}$ and
$\vpt^{\cPgn}=\vPTm$. The utility of \MT\ lies in the fact that $\MT \le
\mw$ is guaranteed for \PW\ bosons with low transverse momentum. For a
single $\PW\to \Pl\cPgn$ decay such a lower limit is only marginally
informative, but in an ensemble of events, the maximum value achieved, \ie
the \textit{endpoint} of the \MT\ distribution, directly reveals the \PW\
boson mass. This observation suggests a ``min-max'' strategy which is
generalized by the invention of \mtt.

The \mtt\ observable is useful for finding the minimum parent mass that is
consistent with observed kinematics when \textit{two} identical decay
chains $a$ and $b$ each terminate in a missing particle.
Figure~\ref{f:topsusy} shows both a SM and a new physics example. If one
knew the two missing transverse momenta separately, a value of \MT\ could
be calculated for either or both of the twin decay chains and the parent
mass $M$ would satisfy the relationship
$\max(\MT^{\text{a}},\MT^{\text{b}}) \le M$. In practice the two missing
momenta cannot be known separately, and are observable only in the
combination $\vpt^{\text{a}}+\vpt^{\text{b}}=\vPTm$. This compels one to
consider all possible partitions of $\vPTm$ into two hypothetical
constituents $\vpt^{\text{a}}$ and $\vpt^{\text{b}}$, evaluating within
this ensemble of partitions the \textit{minimum} parent mass $M$ consistent
with the observed event kinematics. With this extension of the \MT\
concept, the variable is now called \mtt:
\begin{equation}
	\mtt \equiv
	\min_{\vec{p}^{\text{a}}_{\text{T}}+\vec{p}^{\text{b}}_{\text{T}}={\scriptsize\vPTm}}\;
	\left\{\max(M^{\text{a}}_{\text{T}},M^{\text{b}}_{\text{T}})\right\}.
\end{equation}
As with \MT, the endpoint of the \mtt\ distribution has a quantifiable
relationship to the parent mass, and the endpoint of an \mtt\ distribution
is therefore a measure of the unseen parent mass in events with two
identical decay chains.

The observable \mtt\ requires some care in its use. The presence of
$\ET=\sqrt{\pt^{2}+m^{2}}$ in Eq.~\ref{e:mt} implies that one must either
know (as in the case of $\PW\to\Pl\cPgn$) or assume (as in the case of
unknown new physics) a value of the mass $m$ of the undetected particle(s).
In this paper we will refer to an assumed mass as the ``test mass'' and
distinguish it with a tilde (\ie $\widetilde{m}$); the actual mass of the
missing particle, whether known or not, will be referred to as the ``true
mass'', and written without the tilde. Both the value of \mtt\ in any event
and the value of the endpoint of the \mtt\ distribution in an ensemble of
events are in the end {\em functions} of the test mass.

Even when a test mass has been chosen, however, the endpoint of the
\mtt\ distribution may not be unique because it is in general sensitive
to transverse momentum $\Pt=|\vPt|$ of the underlying two-parent system, which
varies from event to event. The sensitivity vanishes if the test mass
can be set equal to the true mass, but such an option will not be immediately
available in a study of new physics where the true mass is not known.

The \Pt\ problem is instead addressed by introducing
\mttp~\cite{Konar:2009wn,Matchev:2009ad}, which uses only momentum
components transverse to the \vPt\ boost direction.  In this way, \mttp\
achieves invariance under \vPt\ boosts of the underlying two-parent
system. The construction of \mttp\ is identical to that of \mtt\ except
that \vpt\ values that appear explicitly or implicitly in Eq.~\ref{e:mt}
are everywhere replaced by \vptp\ values, where \vptp\ is defined to be
the component of \vpt\ in the direction perpendicular to the \vPt\ of
the two-parent system.  Formally,
\begin{equation}
	\label{eq:perp}
	\vptp \equiv \hat{\vec{n}}_{\text{T}}\times
	\left( \vpt \times \hat{\vec{n}}_{\text{T}}\right),
\end{equation}
where $\hat{\vec{n}}_{\text{T}}=\vPt/|\vPt|$ is the unit vector
parallel to the transverse momentum of the two-parent system.

\subsubsection{Subsystem Variables}
A further investigation of \mtt\ and  \mttp\ reveals the full
range of kinematic information contained in multistep decay chains by
splitting and grouping the elements of the decay chain in independent
ways.

The \mtt\ variable classifies the particles in an event into three
categories: ``upstream", ``visible'', and ``child''. The child particles
are those at the end of the decay chain that are unobservable or simply
treated as unobservable; the visible particles are those whose transverse
momenta are measured and used in the calculations; and the upstream
particles are those from further up the decay chain, including any ISR
accompanying the hard collision.

\begin{figure*}[htbp]
    \begin{center}
        \includegraphics[width=0.8\textwidth]{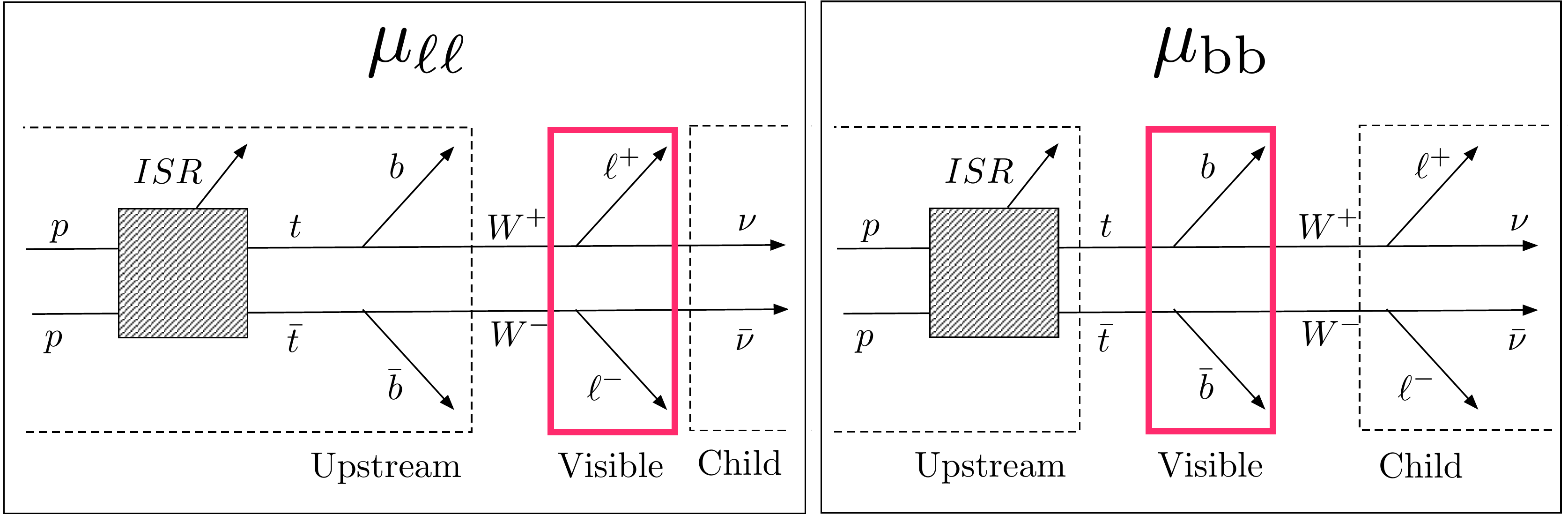}
        \caption{A \ttbar\ dilepton decay with the two subsystems
        for computing \mttpa\ and \mttpb\ indicated.
        The ``upstream'' and ``child'' objects are
        enclosed in dashed rectangles, while the visible objects,
        which enter into the computation, are enclosed in solid rectangles.
        The \mttpa\ and \mttpb\ variables used here are identical to
        \mttpaOLD\ and \mttpbOLD\ of Ref. \citenum{Burns:2008va}.}
        \label{f:ttbardecay}
    \end{center}
\end{figure*}

In general, the child, visible, and upstream objects may actually be
collections of objects, and the subsystem observables introduced in
Ref.~\cite{Burns:2008va} parcel out the kinematic information in as many
independent groupings as possible. Figure~\ref{f:ttbardecay} shows two of
the three possible ways of classifying the \ttbar\ daughters for \mtt\
calculations. The \mttpa\ variable, known as \mttpaOLD\ in Ref.
\citenum{Burns:2008va}, uses the two leptons of the \ttbar\ dilepton decays,
treating the neutrinos as lost child particles (which they are), and
combining the b jets with all other ``upstream'' momentum in the event. The
\mttpb\ variable, known as \mttpbOLD\ in Ref.  \citenum{Burns:2008va}, uses
the b jets, and treats the W bosons as lost child particles (ignoring the
fact that their charged daughter leptons are in fact observable). It
considers only ISR jets as generators of upstream momentum.

For completeness, we note that a third \mttp\ subsystem can be
constructed by combining the b jet and the lepton as a single visible
system. This variable, known as \mttpc\ in the nomenclature of Ref.
\cite{Burns:2008va}, exhibits significant correlation with \mbl, the
invariant mass of the b jet and lepton. A third observable is needed
to solve the underlying system of equations, and for this we choose \mbl.

\subsection{Observables Used in this Analysis}
\label{sec:observables}
This analysis is based on two \mttp\ variables, \mttpa\ and \mttpb\ as
described above, and one invariant mass, \mbl, the invariant mass of a b
jet and lepton from the same top-quark decay. These three quantities have
been selected from a larger set of possibilities based on the low
correlation we observe among them and the generally favorable shapes of the
distributions in their endpoint regions. The observables can be summarized
by the underlying kinematics from which they are derived, and the endpoint
relations which include the top-quark, W-boson, and neutrino masses.

For the \mttpa\ variable, the shape of the distribution is known
analytically \cite{Konar:2009wn}. In terms of the value $x=\mttpa$ and
its kinematic endpoint $x_{\max}$, the normalized distribution can be
written:
\begin{equation}
	\label{e:konstantin}
	\frac{\rd{}N}{\rd{}x} = \alpha\,\delta(x)+(1-\alpha)\,
	\frac{4x}{x_{\max}^2}\ln \frac{x_{\max}}{x},
\end{equation}
where the parameter $\alpha$ is treated as an empirical quantity to be
measured. In practice, $\alpha\sim 0.6$, and the zero bin of \mttpa\
histograms will be suppressed to better show the features of the endpoint
region. The origin of the delta function is geometric: for massless
leptons, $\mttpa$ vanishes when the two lepton \vptp\ vectors lie on
opposite sides of the axis defined by the upstream \vPt\ vector, and
is equal to $2(\ptp^{\ell^+}\, \ptp^{\ell^-})^{1/2}$ otherwise.

For a test mass of the child particle \testmassnu, the endpoint
is related to the masses via \cite{Burns:2008va,Konar:2009wn}:
\ifthenelse{\boolean{cms@external}}{
\begin{multline}
	\label{eq:mttpamax}
	\mttpamax\equiv x_{\max} =\frac{\mw}{2}\left(1-\frac{\mnsq}{\mwsq}\right) +\\
	\sqrt{\frac{\mwsq}{4}\left(1-\frac{\mnsq}{\mwsq}   \right)^2  +\testmassnu^2}.
\end{multline}
}{
\begin{equation}
	\label{eq:mttpamax}
	\mttpamax\equiv x_{\max} =\frac{\mw}{2}\left(1-\frac{\mnsq}{\mwsq}\right) +
	\sqrt{\frac{\mwsq}{4}\left(1-\frac{\mnsq}{\mwsq}   \right)^2  +\testmassnu^2}.
\end{equation}
}
In the \ttbar\ case, we set the test mass to $\testmassnu=0$. We then
expect the endpoint {at $\mttpamax=\mw(1-{\mnsq}/{\mw^2})=\mw=80.4$
\GeV.} Note that \mn\ is the true mass of the child and \mw\ is the true
parent mass; these should be viewed as variables in a function for which
\testmassnu\ is a parameter. In a new-physics application, the analogs of
\mw\ and \mn\ are not known; but given Eq.~\ref{eq:mttpamax}, the
measurement of the endpoint, and an arbitrary choice of child mass
\testmassnu, one can fix a {\em relationship} between the two unknown
masses. We emphasize that the equality expressed by Eq.~\ref{eq:mttpamax}
holds regardless of the value of the test mass, because the test mass enters
into both sides of the equation (see discussion in Section
\ref{sec:mttobservable}).  This applies below to Eq.~\ref{eq:mttpbmax}
also.

In the case of \mttpb, the visible particles are the two b jets, the child
particles are the charged  leptons and neutrinos (combined), and ISR
radiation generates the upstream transverse momentum.  We take the visible
particle masses to be the observed jet masses, which are typically ${\sim}
10$\GeV. The endpoint is unaffected by nonzero jet masses provided the
test mass is set to the true mass, and is affected only at the $\pm$0.1\GeV\
level over a large range of test masses, $0<\testmassW < 2\mw$.
For an assumed child mass \testmassW, the endpoint is given by
\cite{Burns:2008va,Konar:2009wn}:
\begin{equation}
	\label{eq:mttpbmax}
	\mttpbmax=\frac{\mt}{2}\left(1-\frac{\mwsq}{\mtsq}\right) +
	\sqrt{\frac{\mtsq}{4}\left(1-\frac{\mwsq}{\mtsq}   \right)^2  +\testmassW^2}.
\end{equation}
In the \ttbar\ case, we set the test mass to $\testmassW=\mw=80.4$\GeV.
We then expect the endpoint at $\mttpbmax= \mt$. As in the previous case,
in a new-physics application where the analogs of \mt\ and \mw\ are not
known, the measurement of the endpoint together with an arbitrary choice of
the child mass \testmassW\ yields a relationship between the two unknown
masses.

As noted above, a third variable is needed, and we adopt \mbl, the invariant
mass formed out of jet-lepton pairs emerging from the top-quark decay. Two
values of \mbl\ can be computed in a \ttbar\ event, one for each top decay.
In practice four are calculated because one does not know a~priori how to
associate the b jets and leptons; we discuss later an algorithm for
mitigating the combinatorial effects on the endpoint. The shape of the
distribution is known for correct combinations but is not used here since
correct combinations cannot be guaranteed (see Section \ref{s:adam}). The
endpoint is given by:
\begin{equation}
	\label{eq:mblmax}
	\mblmax=\sqrt{
	m_\cPqb^2  +
	\left( 1-\frac{\mnsq}{\mwsq}      \right)
	\left( E^\ast_\PW + p^\ast   \right)
	\left( E^\ast_\cPqb + p^\ast   \right)
	},
\end{equation}
where  $E^\ast_\PW$, $E^\ast_\cPqb$, and $p^\ast$
are energies and momenta of the daughters of $\cPqt\to \cPqb\PW$ in the
top-quark rest frame. In these formulae the charged-lepton mass is
neglected but the observed b-jet mass $m_\cPqb$ is finite and varies
event-to-event.

We can now summarize the mass measurement strategy.  If the masses \mt,
\mw, and \mn\ were unknown, one would measure the two endpoints and the
invariant mass that appear on the left-hand sides of
Eqs.~\ref{eq:mttpamax}--\ref{eq:mblmax}, using arbitrary test mass values
for the first two, to obtain three independent equations for the three
unknown masses. Then, in principle, one solves for the three masses.  In
practice, the measurements carry uncertainties and an optimum solution must
be determined by a fit. In the case when one or more of the masses is
known, a constrained fit can improve the determination of the remaining
unknown mass(es).

\begin{figure*}[htbp]
	\begin{center}
		\includegraphics[width=0.32\textwidth]{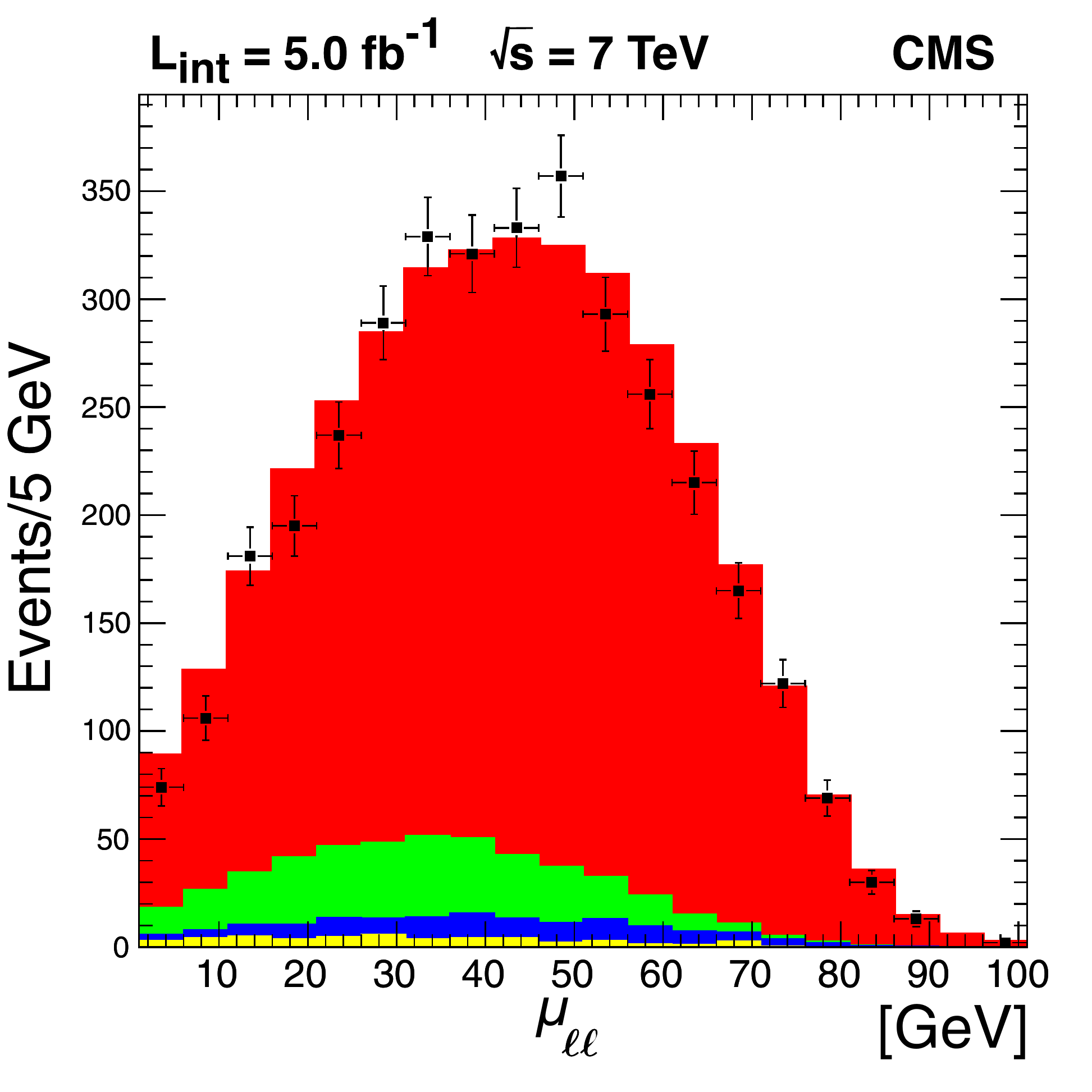}
		\includegraphics[width=0.32\textwidth]{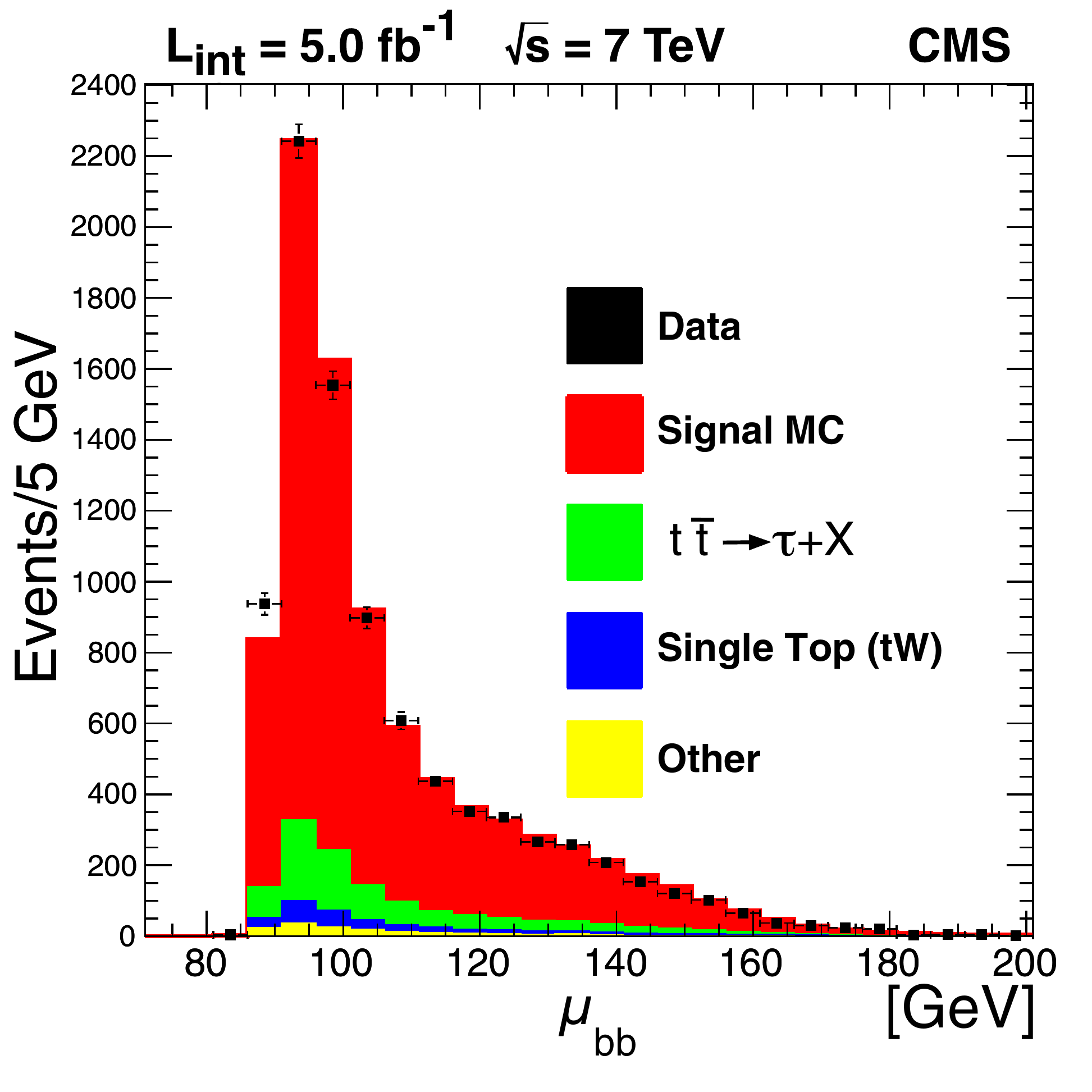}
		\includegraphics[width=0.32\textwidth]{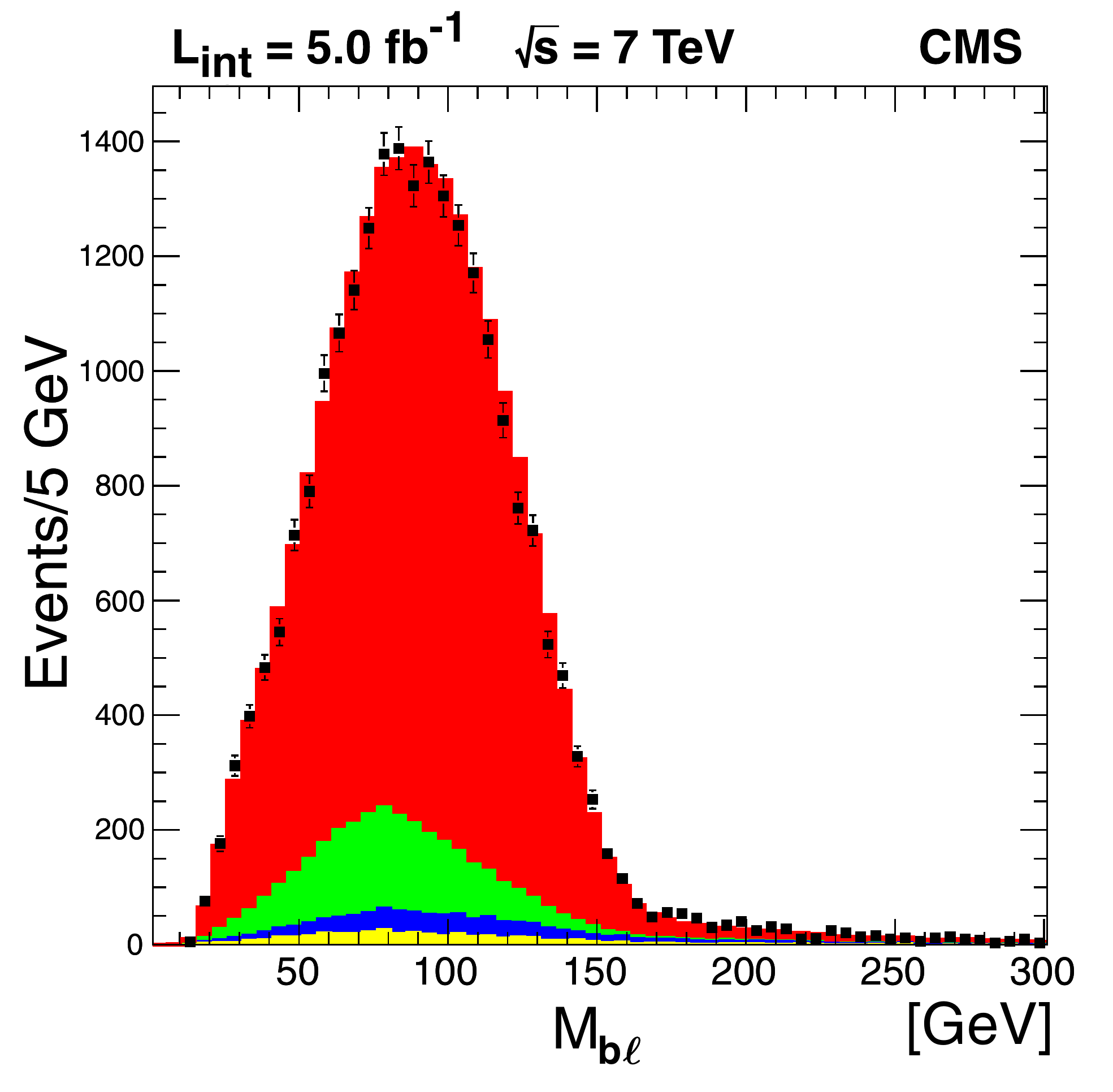}
		\caption{
		Distributions of the three kinematic distributions \mttpa, \mttpb, and
		\mbl. Data (\ourlumi\fbinv) are shown with error bars. MC simulation is
		overlaid in solid color to illustrate the approximate \ttbar\ signal and
		background content of the distributions. The backgrounds contained in
		``Other" are listed in Table~\ref{t:backs}. The zero-bin of the \mttpa\ plot is
		suppressed for clarity. The \mbl\ plot contains multiple entries per event
		(see Section \ref{s:adam} for details). In all cases, the simulation is
		normalized to an integrated luminosity of \ourlumi\fbinv with
		next-to-leading-order (NLO) cross sections as described in the text.
		}
		\label{fig:plots}
	\end{center}
\end{figure*}

In Fig.~\ref{fig:plots} we show distributions for the three observables
\mttpa, \mttpb, and \mbl. Here and throughout this paper, the zero bin of
the \mttpa\ distribution, corresponding to the delta function of
Eq.~\ref{e:konstantin}, is suppressed to emphasize the kinematically
interesting component of the shape. In the \mttpb\ plot shown here, the
prominent peak that dominates the figure is an analog of the delta function
in \mttpa, its substantial width being due to the variable mass of the jets
that enter into the \mttpb\ calculation. As with the \mttpa\ delta
function, the peak arises from events where the axis of the upstream \vPt\
falls between the two visible-object \vpt\ vectors. In later plots this
\mttpb\ peak will be suppressed to better reveal the behavior of the
distribution in the endpoint region.

The agreement between data and MC is generally good, but the comparisons
are for illustration only and the analysis and results that follow do not
depend strongly on the MC simulation or its agreement with observation.

\section{Backgrounds}
\label{sec:backgrounds}

The two-lepton requirement at the core of the event selection ensures an
exceptionally clean sample. Nevertheless a few types of background must be
considered, including top-quark decays with $\tau$-lepton daughters,
$\text{p}\text{p}\to \cPqt \text{W}$ events, and sub-percent contributions
from other sources.

\subsection{Physics Backgrounds}
The physics backgrounds consist of \ttbar\ decays that do not conform to
the dilepton topology of interest, as well as non-\ttbar\ decays. Table~\ref{t:backs} shows the estimation of signal and background events in MC
simulation. The MC generators used throughout this study are
\MCATNLO 3.41~\cite{Frixione:2002ik} for all \ttbar\ samples,
\PYTHIA 6.4~\cite{pythia6.4} for the diboson samples, and
\MADGRAPH 5.1.1.0~\cite{MG5} for all others. The simulated data samples are
normalized to 7\TeV NLO cross sections and an integrated luminosity of
\ourlumi\fbinv.

Events in which a top quark decays through a $\tau$ lepton
(\eg\ $\cPqt\to \cPqb\tau^+\nu_{\tau}\to \cPqb \Pl^+\nu_{\Pl}\bar\nu_{\tau}\nu_{\tau}$),
constitute about 13\% of the events surviving all selection requirements.
From the point of view of event selection, these events are background. The
unobserved momentum carried by the extra neutrinos, however, ensures that
these events reconstruct to \mtt\ and \mbl\ values below their true values
and hence fall below the endpoint of signal events with direct decays to
$\Pe$ or $\Pgm$ final states. We therefore include these events among the
signal sample. This leaves in principle a small distortion to the kinematic
shapes, but the distortion is far from the endpoint and its impact on the
mass extraction is negligible.

\begin{table}[htbp]
  \topcaption{
  Estimate of signal and background composition in MC simulation,
  normalized to an integrated luminosity of \ourlumi\fbinv and NLO cross
  sections as described in the text.
  }
  \begin{center}
    \begin{tabular}{lr}
      \hline
      Process                                               &  Number of Events\cr\hline
      \ttbar\ signal (no $\Pgt$)                            & 7000 \cr
      \ttbar\ signal ($\Pgt\to \Pl\cPgn$)                   & 1100 \cr
      \hline
      Single top ($\cPqt\PW,\bar{\cPqt}\PW$)                            & 270 \cr
      Drell--Yan                                             & 77 \cr
      Hadronic/Semileptonic $\ttbar$     & 55 \cr
      \qquad with misreconstructed lepton(s) & \cr
      Dibosons (WW, ZZ, WZ)                                   & 14 \cr
      W+jets                                                & 9  \cr
      \hline
    \end{tabular}
  \end{center}
  \label{t:backs}
\end{table}%

\subsection{Modelling the Mistag Background}
\label{s:kde}

\begin{figure}[htbp]
	\begin{center}
		\includegraphics[width=0.48\textwidth]{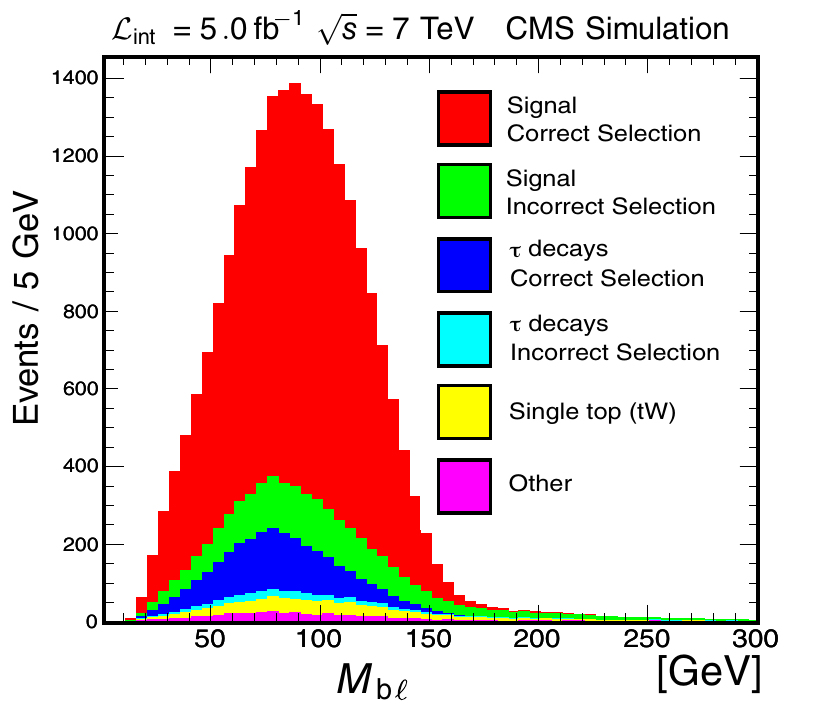}
		\includegraphics[width=0.48\textwidth]{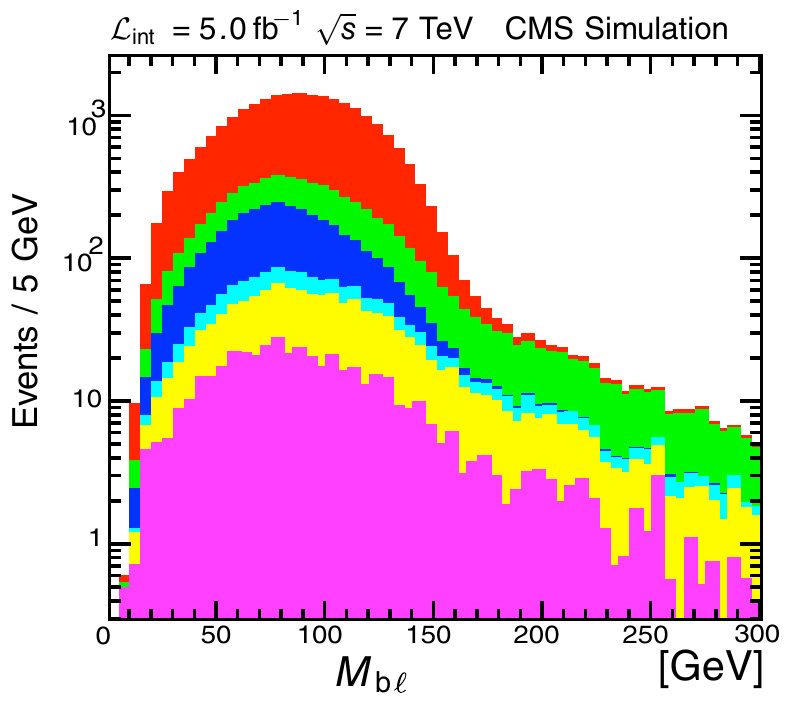}
		\caption{
		Composition of MC event samples, illustrating that signal events
		with light-quark and gluon jet contamination dominate the region
		beyond the endpoint. The \cmsLeft and \cmsRight \mbl\ distributions contain
		the same information plotted with different vertical scales.  The
		backgrounds contained in ``Other" are listed in Table~\ref{t:backs}.
		}
		\label{f:beyond}
	\end{center}
\end{figure}

In addition to the backgrounds discussed above, which fall within the bulk
the distributions, it is essential also to treat events that lie beyond the
nominal signal endpoint.  In this analysis, the main source of such events
comes from genuine \ttbar\ events where one of the jets not originating from
a top-quark decay is mistagged as a b jet. An event in which a light-quark
or gluon jet is treated as coming from a top quark can result in events
beyond the endpoint in the \mttpb\ and \mbl\ distributions, as can be seen
in Fig.~\ref{f:beyond}.  The measurement of \mttpa, on the other hand,
depends primarily on the two leptons and is unaffected by mistags.

To determine the shape of the mistag background in \mttpb\ and \mbl, we
select a control sample with one b-tagged jet and one antitagged jet, where
the antitagging identifies jets that are more likely to be light-quark or
gluon jets than b jets. Antitagging uses the same algorithm as combined
secondary vertex algorithm, but selects jets with a low discriminator value
to obtain a sample dominated by light-quark and gluon jets. We classify
event samples by the b-tag values of the two selected jets, and identify
three samples of interest: a signal sample where both jets are b-tagged; a
background sample where one jet is b-tagged and the other antitagged; and
another background sample where both jets are antitagged.
Table~\ref{tab:comp} shows the composition of these samples as determined
in MC simulation. We select the sample consisting of pairs with one tagged
and one antitagged jet to be the control sample and use it to determine the
shape of the background lying beyond the signal endpoint. It contains a
significant fraction of signal events, 27\%, but these all lie below the
endpoint and categorizing them as background does not change the endpoint
fit.

\begin{table}[htbp]
  \topcaption{
  Composition of b-tagged, dijet samples as determined in MC simulation.
  Each column is an independently selected sample; columns sum to 100\%.
  }
  \begin{center}
     \begin{tabular}{rccc}
      \hline
      &  2 b-tags   & b-tag, antitag    &  2 antitags\\ \hline
      b jet, b jet   & $86\%$ & $27\%$ & $7.1\%$ \\
      b jet, non b jet  & $14\%$ & $70\%$ & $53\%$ \\
      non b jet, non b jet   & $0.3\%$ & $3\%$ & $40\%$ \\
      \hline
    \end{tabular}
  \end{center}
  \label{tab:comp}
\end{table}%

\begin{figure}[htbp]
	\begin{center}
	\includegraphics[width=0.48\textwidth]{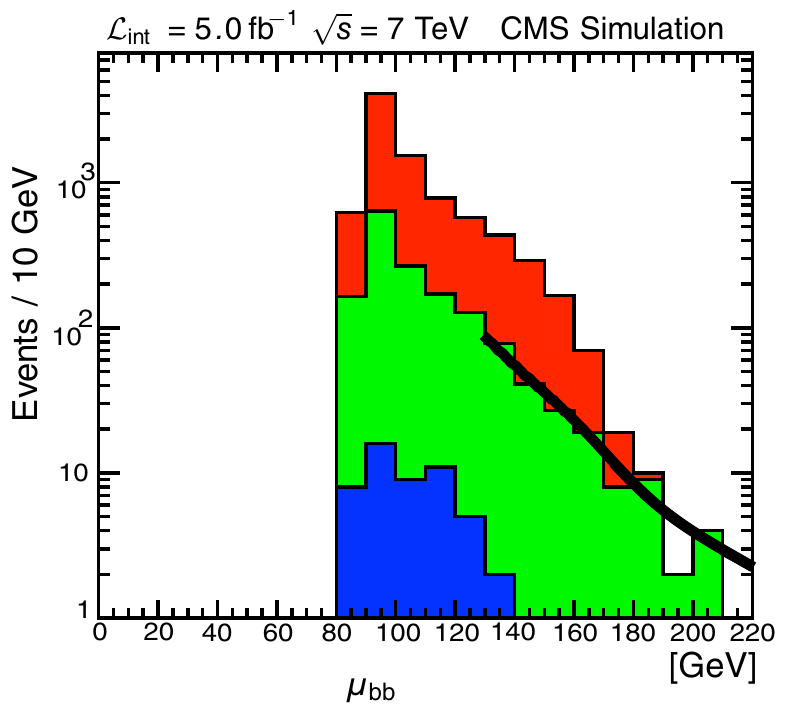}
	\includegraphics[width=0.48\textwidth]{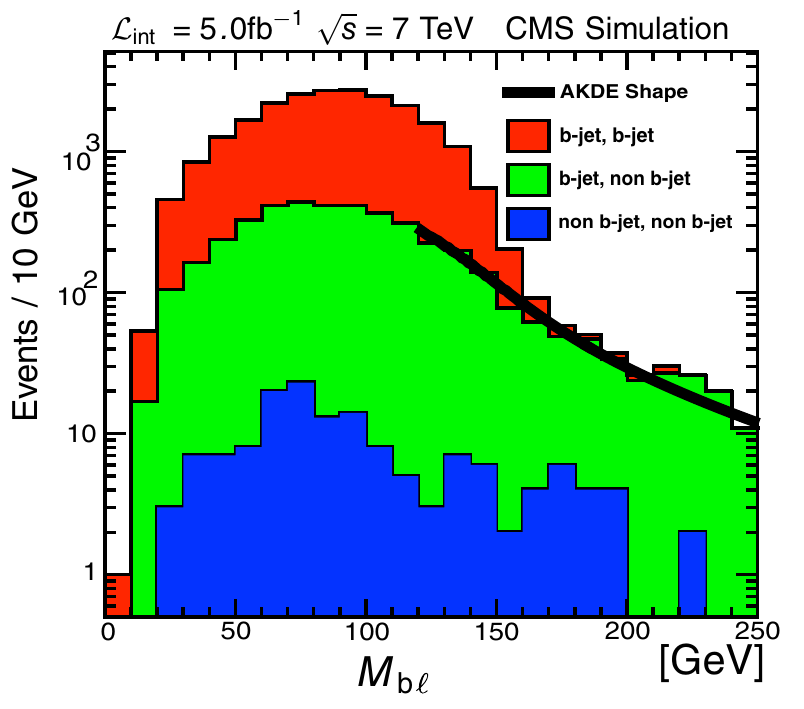}
	\caption{
	Background PDF shapes determined by the AKDE method, on MC samples. All
	events pass the signal selection criteria. \cmsLeft: \mbl; \cmsRight: \mttpb.
	The heavy black curve is the AKDE shape.}
	\label{f:kde}
	\end{center}
\end{figure}

The control sample is used to generate distributions in \mttpb\ and \mbl,
whose shapes are then characterized with an adaptive kernel density
estimation (AKDE) method \cite{DensityEstimation}.  The underlying KDE
method is a non-parametric shape characterization that uses the actual
control sample to estimate the probability distribution function (PDF) for
the background by summing event-by-event Gaussian kernels. In the AKDE
algorithm, on the other hand, the Gaussian widths depend on the local
density of events; empirically this algorithm yields lower bias in the
final mass determination than alternative algorithms. Figure~\ref{f:kde}
shows the performance of the background shape determination; the set of
control sample events are taken from MC simulation in order to illustrate
the composition of the background and signal.

\subsection{Suppressing the Combinatorial Background}
\label{s:adam}
Even if the b-tagging algorithm selected only b jets, there would remain a
combinatorics problem in \ttbar\ dilepton events. In the case of the \mbl\
distribution the matching problem arises in pairing the b jet to the
lepton:  for b jets $j_1$ and $j_2$, and leptons $\ell^+$ and $\Pl^-$,  two
pairings are possible: $j_1\Pl^+,j_2\Pl^-$ and $j_1\Pl^-,j_2\Pl^+$. Four
values of \mbl\ will thus be available in every event, but only two of them
are correct. The two incorrect pairings can (but do not have to) generate
values of \mbl\ beyond the kinematic endpoint of \mbl\ in top-quark decay.
To minimize the unwanted background of incorrect pairings while maximizing
the chance of retaining the highest values of \mbl\ in correct b$\Pl$
pairings, which do respect the endpoint, we employ the following algorithm.

Let $A$ and $a$ denote the two \mbl\ values calculated from one of the
two possible b$\Pl$ pairings, and let $B$ and $b$ denote the \mbl\
values calculated from the other pairing. Choose the labeling  such
that $a<A$ and $b<B$. Without making any assumptions about which pairing
is correct, one can order the \mbl\ values from smallest to largest;
there are six possible orderings. For example the ordering $b,B,a,A$
means that the $bB$ pairing has \mbl\ values which are both smaller than
the \mbl\ values in the $aA$ pairing.  In this case, while we do not
know which pairing is {\em correct}, we can be certain that both \mbl\
values of the $bB$ pairing must respect the true endpoint since either
(a) $bB$ is a correct pairing, in which case its \mbl\ values naturally
lie below the endpoint, or (b) $aA$ is the correct pairing, so its \mbl\
values lie below the true endpoint, with the $bB$ values falling at yet
lower values.  Similar arguments apply to each of the other possible orderings.

Table~\ref{t:mblorderings} shows the six possibilities. For each mass ordering
shown in the left column, the right column shows the mass values that will be
selected for use in the \mbl\ fit. For any given event only one row of the table
applies. For an event falling in one of the first two rows, two values of \mbl\
enter in the subsequent fits; for an event falling in the last four rows, three
values enter the fits.

\begin{table}[htbp]
  \topcaption{
  \mbl\ orderings: in each column the left-to-right sequencing of the
  $a,A,b,B$ labels is from lowest \mbl\ value to highest. The left column
  lists the six possible \mbl\ orderings; the right column indicates for
  each ordering which values are selected for inclusion in the \mbl\ plot.}
  \begin{center}
    \begin{tabular}{ll}
      \hline
      Ordering        &  Selection \cr
      \hline
      $bBaA$          & $b,B$ \cr
      $aAbB$          & $a,A$ \cr
      $baBA$          & $b,a,B$ \cr
      $baAB$          & $b,a,A$ \cr
      $abBA$          & $a,b,B$ \cr
      $abAB $         & $a,b,A$ \cr
      \hline
    \end{tabular}
  \end{center}
  \label{t:mblorderings}
\end{table}

This selection algorithm ensures that all masses used in the fits that can
be guaranteed to be below the endpoint will be used, while any that
\textit{could} exceed the endpoint because of wrong pairings will be
ignored. Note that it does not guarantee that the masses that are used are
all from correct b$\Pl$ pairings; in practice, however, we find that 83\%
of the entries in the fit region are correct b$\Pl$ pairings, and that this
fraction rises to over 90\% within 10\GeV of the endpoint.

\section{Fit Strategy}

The kinematic observables \mttpa, \mttpb, and \mbl, along with their
endpoint relations (Section \ref{sec:observables}) and background
mitigation techniques (Sections \ref{s:kde}, \ref{s:adam}), are combined in
an unbinned event-by-event maximum likelihood fit. The likelihood function
is given by a product over all events of individual event likelihoods
defined on each of the kinematic variables:
\begin{equation}
\mathcal{L}(\vec{M})=\prod_{i=1}^{N}
				{\mathcal L}_i^{\mttpa}(\vec{u}_i|\vec{M})\cdot
				{\mathcal L}_i^{\mttpb}(\vec{u}_i|\vec{M})\cdot
				{\mathcal L}_i^{\mbl}(\vec{u}_i|\vec{M}).
\label{eq:like}
\end{equation}
The vector $\vec{M}=(\mt,\mw,\mnsq)$ contains the mass parameters to be
determined by the fit, and each $\vec{u}_{i}$ comprises the set of
transverse momentum vectors, reconstructed object masses, and
missing-particle test masses from which the kinematic observables \mttpa,
\mttpb, and \mbl\ of the event $i$ are computed. We fit for \mnsq\ rather
than \mn\ because only \mnsq\ appears in the endpoint formulae
(Eqs.~\ref{eq:mttpamax} and \ref{eq:mblmax}); we do not constrain \mnsq\  to
be positive.   As will be described more fully below, only the endpoint
region of each variable is used in the fit. If an event $i$ does not fall
within the endpoint region of a given variable, the corresponding
likelihood component ($\mathcal{L}_i^{\mttpa}$, $\mathcal{L}_i^{\mttpb}$, or
$\mathcal{L}_i^{\mbl}$) defaults to unity.

For each observable $x\in \{\mttpa, \mttpb, \mbl \}$, the  likelihood
component ${\mathcal L}_i$ in Eq.~\ref{eq:like} can be expressed in terms
of the value of the observable itself, $x_i=x(\vec{u}_i)$, and its
kinematic endpoint, $x_{\max}=x_{\max}(\mathbf{M})$. Explicit formulae for
$x_{\max}(\mathbf{M})$ are given in Eqs.~\ref{eq:mttpamax},
\ref{eq:mttpbmax}, and \ref{eq:mblmax}; in the first two cases there is
additional dependence on the missing-particle test mass. Letting the label
$\text{a}\in\{\ell\ell, \text{b}\text{b}, \bell \}$ index the three flavors
of observables, we can write the signal, background, and resolution shapes
as $S(x|x^\text{a}_{\max})$, $B^{\text{a}}(x)$,  and $\mathcal
R^{\text{a}}_i(x)$. While the form of the signal shape $S(x)$ is common to
all three fits, the background shape $B^{\text{a}}(x)$ is specific to each
observable and the resolution function $\mathcal R^{\text{a}}_i(x)$ is
specific to both the observable and the individual event. Then each
function $\mathcal L^{\text{a}}_i$ appearing on the right-hand side of
Eq.~\ref{eq:like} is given by the general form:
\begin{equation}
	\label{eq:smalllike}
	{\mathcal L}^{\text{a}}_i(x_i|x^\text{a}_{\max}) = \beta\! \int \!
	\! S(y|x^\text{a}_{\max}) {\cal R}^{\text{a}}_i(x_i-y)\,\rd{}y
	+ (1-\beta) B^{\text{a}}(x_i).
\end{equation}
The fit parameter $\beta$ determines the relative contribution of
signal and mistag background.

For the common signal shape $S(x|x^\text{a}_{\max})$ we use an
approximation consisting of a kinked-line shape,  constructed piecewise
from a descending straight line in the region just below the endpoint and a
constant zero value above the endpoint. The kinked-line function is defined
over a range from $x_{\text{lo}}$ to $x_{\text{hi}}$. The generic form is:
\begin{equation}
	S(x|x_{\max})\equiv
	\left\{
	\begin{array}{ll}
	\mathcal{N}
	(x_{\max} -x) & x_{\text{lo}}  \le x \le x_{\max}; \cr
	0    & x_{\max}\le x \le x_{\text{hi}}.
	\end{array}
	\right.
	\label{e:f221}
\end{equation}

The parameter ${\cal N}$ is fixed by normalization. The fidelity of this
first-order approximation to the underlying shape depends on both the shape
and the value of $x_{\text{lo}}$. The range of the fit,
$(x_{\text{lo}},x_{\text{hi}})$, is chosen to minimize the dependence of
the fit results on the range, and then the values of $x_{\text{lo}}$ and
$x_{\text{hi}}$ are subsequently varied to estimate the corresponding
systematic uncertainties.

The following paragraphs discuss the forms of $B^{\text{a}}(x)$ and
${\mathcal R}^{\text{a}}(x)$ for each of the three kinematic distributions.

\subsection{\texorpdfstring{\mttpa}{} }
In the case of \mttpa, the visible particles are the two leptons, which are
well measured. The projection of their vectors onto the axis orthogonal to
the upstream \vPt, however, necessarily involves the direction of the
upstream \vPt, which is not nearly as well determined. The resolution
function is therefore wholly dominated by the angular uncertainty in \vPt,
and it varies substantially from event to event depending on the particular
configuration of jets found in each event. Although jet resolutions are
known to have small non-Gaussian tails, their impact on the \mttpa\
resolution function and the subsequent fit procedure is small and we treat
only the Gaussian core. A far more important feature of the resolution
arises when the \vPt\ direction uncertainty is propagated into the \mttpa\
variable to derive ${\mathcal R}^{\ellell}_i(x)$. In this procedure a sharp
Jacobian peak appears wherever the \vPt\ smearing can cause \mttpa\ to pass
through a local maximum or minimum value. These peaks depend only on
azimuthal angles and occur at any value of \mttpa. The detailed shape of
the highly non-Gaussian \mttpa\ resolution and its convolution with the
underlying signal shape, as specified in Eq.~\ref{eq:smalllike}, are
handled by exact formulae derived analytically (see Appendix). The
background in the \mttpa\ distribution is vanishingly small, so we set
$B^{\ellell}(x)=0$.

\subsection{\texorpdfstring{\mttpb}{}{} }
For \mttpb, the visible particles are the  b jets, and since the resolution
smearing of both the b jets and the upstream jets defining \vPt\ are large
and of comparable magnitudes, the event-by-event resolution is more
complicated than in the \mttpa\ case.  As a result, no analytic calculation
is possible and we instead determine the \mttpb\ resolution function,
${\cal R}^{\beebee}_i(x)$,  numerically in each event, using the known
$\pt$ and $\phi$ resolution functions for the jets. As with the \mttpa\
resolutions, Jacobian peaks appear in the \mttpb\ resolutions. The mistag
background is included by scaling the shape $B^{\beebee}(x)$ obtained from
the AKDE procedure as discussed in Section \ref{s:kde}.

\subsection{\texorpdfstring{\mbl}{}}
In the \mbl\ case, the theoretical shape $S(x)$ is well-known, but the
combinatorics of $\cPqb\Pl$ matching, together with the method of selecting
$\cPqb\Pl$ pairs from the available choices (see Section \ref{s:adam}),
sculpt the distribution to the degree that the theoretical shape is no
longer useful. Therefore we use the kinked-line shape of Eq.~\ref{e:f221}
to model the signal near the endpoint. In contrast to the \mttpa\ and
\mttpb\ variables, numerical studies confirm that linearly propagated
Gaussian resolutions accurately reflect the smearing ${\mathcal
R}^{\bell}_i(x)$ of \mbl, as one expects in this case.  The background
shape $B^{\bell}(x)$ is given by the AKDE procedure as discussed in
Section~\ref{s:kde}.

\subsection{Applying the Fit to Data}
\label{s:apply}
The unbinned likelihood fit prescribed in Eqs.~\ref{eq:like} and
\ref{eq:smalllike} is performed on the three kinematic distributions
using the shapes given for signal $S(x|x_{\max})$, resolution ${\mathcal
R}_i(x)$, and mistag background $B(x)$. Although a simultaneous fit for all
three masses is an important goal of this study, it is useful in the
context of the \ttbar\  data sample to explore subclasses of the fit in
which some masses are constrained to their known values. For this purpose
we define: (a) the {\em unconstrained} fit, in which all three masses are
fit simultaneously; (b) the \textit{singly-constrained} fit, in which $\mn=0$
is imposed; and (c) the \textit{doubly-constrained} fit, in which both $\mn=0$
and $\mw=80.4$\GeV are imposed \cite{Beringer:1900zz}. The unconstrained
fit is well-suited to testing mass measurement techniques for new physics,
while the doubly-constrained fit is optimal for a SM determination of the
top-quark mass.

The fit procedure takes advantage of bootstrapping techniques
\cite{EfronBook}. In particle physics, bootstrapping is typically
encountered in situations involving limited MC samples, but it can be
profitably applied to a single data sample, as in this analysis. The goal
of bootstrapping is to obtain the sampling distribution of a statistic of a
particular data set from the data set itself. With the distribution in
hand, related quantities such as the mean and variance of the statistic are
readily computable.

In order to estimate the sampling distribution of a statistic, we first
need to estimate the distribution from which the data set was drawn. The
basic assumption of bootstrapping is that the best estimate for this
distribution is given by a normalized sum of delta functions, one for each
data point. This is the bootstrap distribution. One can estimate the
distribution of a statistic of the data by drawing samples from the
bootstrap distribution and calculating the statistic on each sample. To
simplify the process further we note that, since the bootstrap distribution
is composed of a delta function at each data point, sampling from the
bootstrap distribution is equivalent to sampling from the observed data.

\begin{figure}[htbp]
	\begin{center}
		\includegraphics[width=\cmsFigWidth]{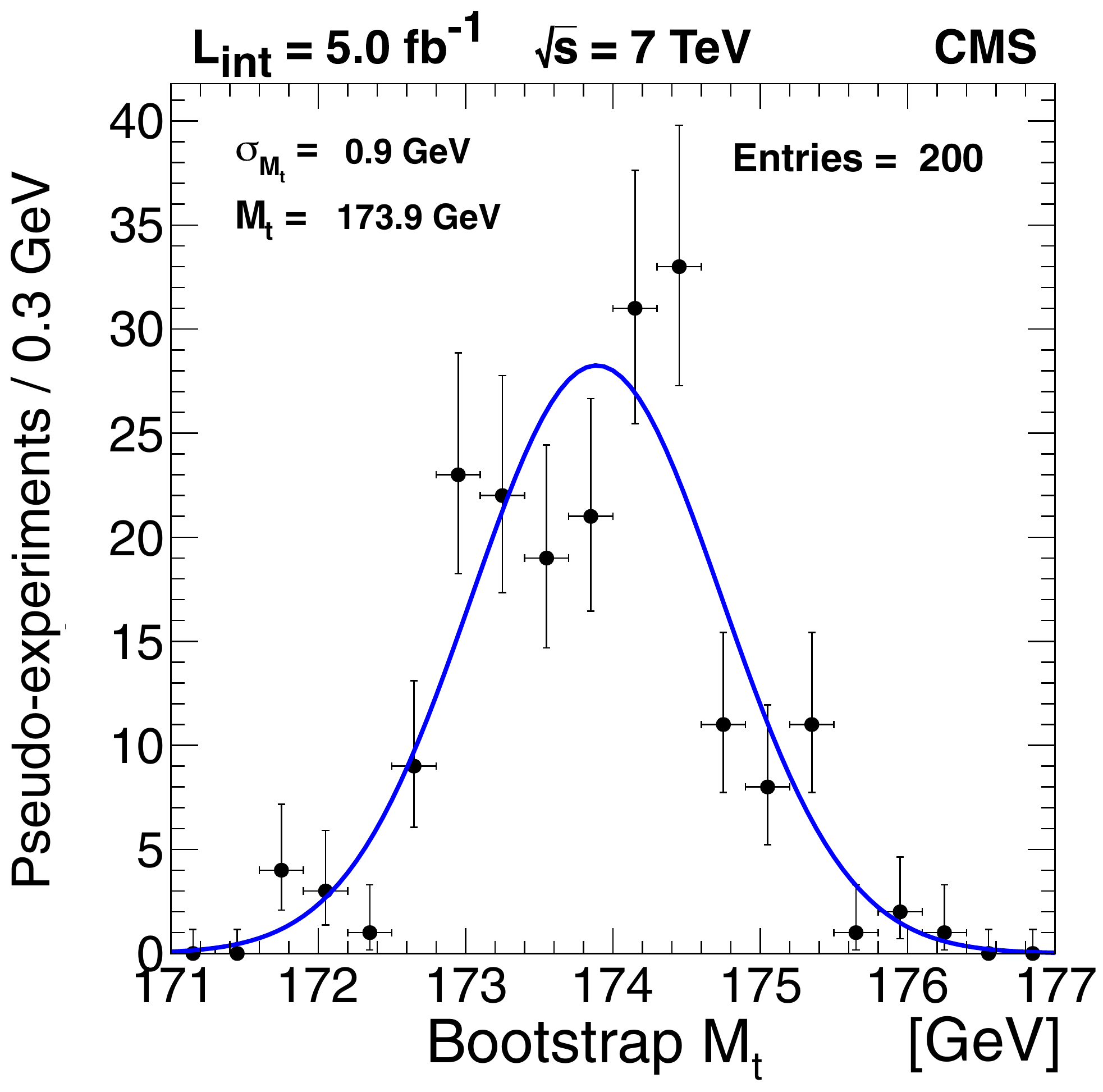}
		\caption{
		Distribution of \mt\ in doubly-constrained fits of 200
		pseudo-experiments bootstrapped from the full data set.}
		\label{f:boot}
	\end{center}
\end{figure}

In this analysis, the fitted top-quark mass is the statistic of interest,
and we wish to find its mean and standard deviation. To do so, we conduct
the fit 200 times, each time extracting a new sampling of events from the
8700 selected events in the signal region of the full data set. The
sampling is done \textit{with replacement}, which means that each of these
bootstrapped pseudo-experiments has the same number of events ($N=8700$) as
the original data set, and any given event may appear in the bootstrap
sample more than once. Each bootstrapped sample is fit with the unbinned
likelihood method described in the previous subsections.  As an
illustration, we show in Fig.~\ref{f:boot} the distribution of the 200
values of \mt\ that emerges in the case of the doubly-constrained fit; the
mean and its standard deviation in this distribution, $\mt = 173.9 \pm
0.9\GeV$, constitute the final result of the doubly-constrained fit.

A key motivation for applying bootstrapping to the data is that the impact
of possible fluctuations in the background shape are naturally
incorporated. Because the background shape in a given fit is constructed
from a control sample with the AKDE method (Section \ref{s:kde}), the
possible statistical variation in the shape is most easily accounted for by
multiple samplings of the control sample. Thus for each bootstrap sample
taken from the signal region of the data, another is taken simultaneously
from the set of background control events. Each pseudo-experiment fit
therefore has its own background function and the ensemble of all 200 such
fits automatically includes background shape uncertainties. (The total
background yield is a separate issue, handled by the normalization
parameter that floats in each fit.)

A secondary motivation to use bootstrapping on the data sample is that it
offers a convenient mechanism to correct for event selection and
reconstruction efficiencies~\cite{Canty}. To do so, each event is assigned
a sampling weight equal to the inverse of its efficiency, and during the
bootstrap process events are selected with probabilities proportional to
these weights.  A bootstrapped data set therefore looks like
efficiency-corrected data, but each event is whole and unweighted.

\section{Validation}
\label{sec:val}

{\tolerance=500 
We test for bias in the above procedures by performing pseudo-experiments
on simulated events. Each pseudo-experiment yields a measurement and its
uncertainty for \mt. From these a pull can be calculated, defined by
$\text{pull} =
({m_{\text{meas}}-m_{\text{gen}}})/{\sigma_{\text{meas}}}\text{.}$ In this
expression $m_{\text{gen}}$ is the top-quark mass used in generating events
while $m_{\text{meas}}$ and $\sigma_{\text{meas}}$ are the fitted mass and
its uncertainty, determined for each pseudo-experiment. For an unbiased
fit, the pull distribution will be a Gaussian of unit width and zero mean.
A non-zero mean indicates the method is biased, while a non-unit width
indicates that the uncertainty is over- or under-estimated. We increase the
precision with which we determine the pull distribution width by
bootstrapping the simulation to generate multiple pseudo-experiments. The
results of Ref. \cite{Barlow} are then used to calculate the mean and
standard deviation of the pull distribution, along with uncertainties on
each, taking into account the correlations between pseudo-experiments
introduced by over-sampling.
\par}

\begin{figure}[ht]
	\begin{center}
	\includegraphics[width=\cmsFigWidth]{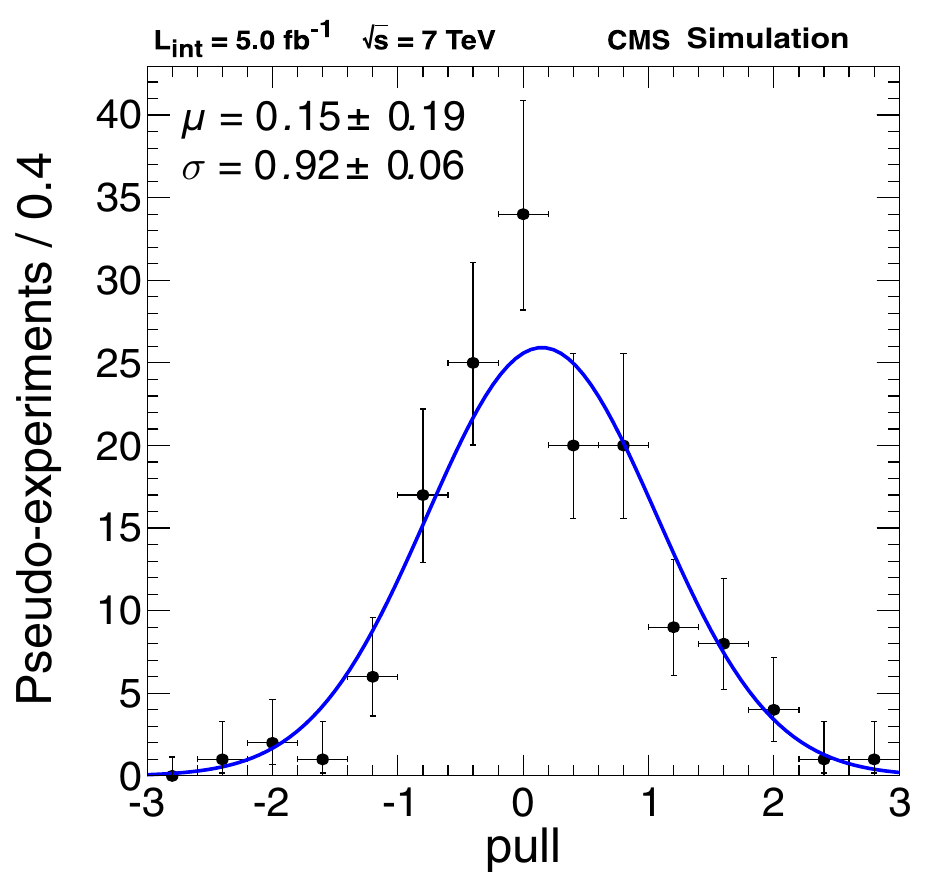}
	\includegraphics[width=\cmsFigWidth]{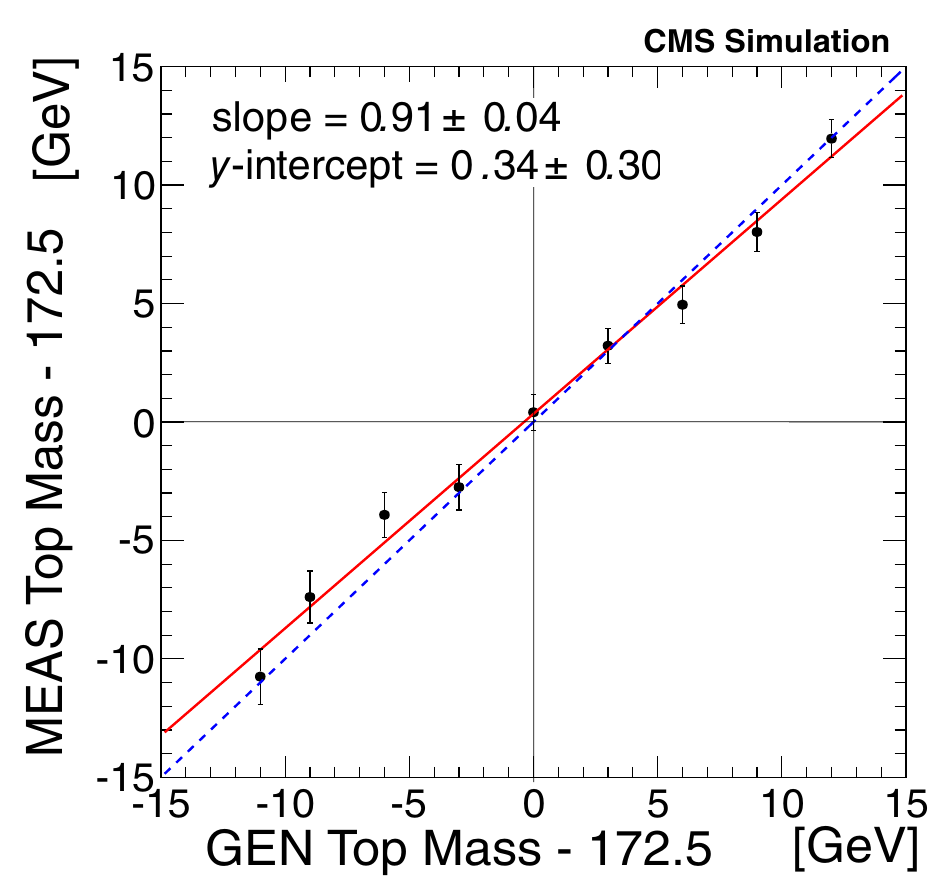}
	\end{center}
	\caption{
	(\cmsLeft) Pull distribution
	${(m_{\text{meas}}-m_{\text{gen}})}/{\sigma_{\text{meas}}}$ for the
	top-quark mass (other masses are fixed) across 150 MC
	pseudo-experiments. (\cmsRight) Fit results obtained in MC \ttbar-only
	samples generated with \MADGRAPH\ for various top-quark masses. The
	best-fit calibration is shown by the solid line and the line of unit
	slope is shown in the dashed line. Data points are from
	doubly-constrained fits. The line of unit slope agrees with the fit
	results with $\chi^2/{\textrm{degree of freedom}=}  10.7/9$. }
	\label{f:valid}
\end{figure}

Figure~\ref{f:valid}, \cmsLeft, shows the pull distribution for the
doubly-constrained fit over 150 pseudo-experiments. Extracting a result
from each pseudo-experiment involves the methods discussed in Section
\ref{s:apply}, and thus the total number of pseudo-experiments required for
the study is $150\times 200$. The measured pull mean is $0.15 \pm 0.19$ and
the pull standard deviation is $0.92 \pm 0.06$, indicating that the fit is
unbiased to the level at which it can be measured with the available
simulated data. The slightly low standard deviation suggests that the
statistical uncertainty may be overestimated; since the systematic
uncertainty is significantly larger than the statistical error, we do not
make any correction for this.

In an independent test, we perform fits to MC samples generated
with various \mt\ values. As the results, shown in Fig.~\ref{f:valid},
\cmsRight, indicate that there is no significant bias as a function of the
top-quark mass, we make no correction.

\section{Systematic Uncertainties}\label{sec:syst}

The systematic uncertainties are assessed by varying the relevant aspects
of the fit and re-evaluating the result.  All experimental systematic
uncertainties are estimated in data. In the doubly-constrained fit,
uncertainties are evaluated for the fitted top-quark mass \mt.

The systematic uncertainties related to absolute jet energy scale (JES) are
derived from the calibration outlined in Ref.~\cite{Chatrchyan:2011ds}. We
evaluate the effects of JES uncertainties in this analysis by performing
the analysis two additional times: once with the jet energies increased by
one standard deviation of the JES, and once with them decreased by the same
amount. Each jet is varied by its own JES uncertainty, which varies with
the  \pt\ and $\eta$ values of the jet. In a generic sample of multijet
events, selecting jets above 30\GeV, the fractional uncertainty in the JES
(averaged over $\eta$) ranges from 2.8\% at the low end to 1\% at high \pt.
The uncertainty is narrowed further by using flavor-specific corrections to
b jets. A similar process is carried out for varying the jet energy
resolutions by its uncertainties.  These variations of jet energy scale and
jet energy resolution propagate into uncertainties of ${}^{+1.3}_{-1.8}$\GeV
and $\pm$0.5\GeV on the measured top-quark mass, respectively.
For the electrons, the absolute energy scale is known to 0.5\% in the barrel
region and 1.5\% in the endcap region, while for the muons the
uncertainty is 0.2\% throughout the sensitive volume.
Varying the lepton energy scale accordingly
leads to a systematic uncertainty in \mt\ of ${}^{+0.3}_{-0.4}\GeV$.

The choice of fit range in \mttpb\ and \mbl\ introduces an uncertainty due
to slight deviations from linearity in the descending portion of these
distributions.  Separately varying the upper and lower ends of the \mttpb\
and \mbl\ fit range gives an estimate of $\pm$0.6\GeV for the systematic
uncertainty.  The uncertainty is mainly driven by dependence on the lower
end of the \mttpb\ range.   A cross-check study based on the methods of
Ref. \cite{PhysRevD.85.075004} confirms the estimate.

The AKDE shape which is used to model the mistag background in \mttpb\ and
\mbl\ is non-parametric and derived from data.  For this reason, the AKDE
is not subject to biases stemming from assumptions about the underlying
background shape or those inherent in MC simulation.  However, one could
also model the mistag background with a parametric shape, and we use this
alternative as a way to estimate the uncertainty due to background
modeling.  Based on comparisons among the default AKDE background shape and
several parametric alternatives, we assign a systematic uncertainty of $\pm$
0.5\GeV.

Efficiency can affect the results of this analysis if it varies across the
region of the endpoint in one or more of the kinematic plots.  The \mbl\
observable is sensitive to both b-tagging and lepton efficiency variations,
whereas \mttpb\ is only sensitive to uncertainties due to b-tagging
efficiency.  By varying the b-tagging and lepton selection efficiencies
by ${\pm}1\sigma$, including their variation with \pt, we
estimate that the effect of the efficiency uncertainty contributes at most
$^{+0.1}_{-0.2}$\GeV uncertainty to the measured top-quark mass.

The dependence on pileup is estimated by conducting studies of fit
performance and results with data samples that have been separated into
low-, medium-, and high-pileup subsamples of equal population; these
correspond to 2--5, 6--8, and $\ge$9 vertices, respectively. The
dependence is found to be negligible.  In addition, direct examination of
the variables \mttpb\ and \mbl\ reveals that their correlation with the
number of primary vertices is small, with correlation coefficients $< 3\%$
and $< 1\%$, respectively.

The sensitivity of the result to uncertainties in QCD calculations is
evaluated by generating simulated event samples with varied levels of
color-reconnection to beam remnants, renormalization and factorization scale,
and jet-parton
matching scale. The impact of the variations on \mt\ is dominated by the
color reconnection effects, which are estimated by comparing the results of
simulations performed with two different MC tunes~\cite{Skands:2010ak},
Perugia2011 and Perugia2011noCR. Factor-of-two variations of
renormalization and factorization scale and the jet-parton
matching scale translate to negligible ($<$0.1\GeV) variations in the top-quark
mass. Uncertainties in the parton distribution functions and relative fractions
of different production mechanisms do not affect this analysis.
The overall
systematic error attributed to QCD uncertainties is ${\pm}0.6\GeV$ on the
value of \mt. In quadrature with other systematic uncertainties these
simulation-dependent estimates add 0.1\GeV to both the upper and lower
systematic uncertainties. This additional contribution reflects theoretical
uncertainty in the interpretation of the measurement as a top-quark mass,
and unlike other systematic uncertainties in the measurement, is
essentially dependent on the reliability of the MC modeling.

For the unconstrained and singly-constrained fits, where the objective is
primarily to demonstrate a method, rather than to achieve a precise result,
we have limited the investigation of systematic uncertainties to just the
evaluation of the jet energy scale and fit range variations, which are
known from the doubly-constrained case to be the dominant systematic
contributions.   Because of this, the systematic uncertainties displayed
for these fits are slightly lower than they would be with a fuller
treatment of all contributions.

The systematic uncertainties discussed in this section are summarized in
Table \ref{t:systematics}.

\begin{table}[htdp]
	\topcaption{
	Summary of systematic uncertainties $\delta \mt$
	affecting the top-quark mass measurement; see text for discussion.
	}
	\begin{center}
	{\renewcommand{\arraystretch}{1.2}
	\begin{tabular}{lc}
	\hline
	~~~Source~~~    & ~~~ $\delta\mt$ (\GeV)~~~ \cr\hline	
	Jet Energy Scale & ${}^{+1.3}_{-1.8}$\cr	
	Jet Energy Resolution &  ${\pm}0.5$\cr	
	Lepton Energy Scale & ${}^{+0.3}_{-0.4}$\cr		
	Fit Range &   ${\pm}0.6$\cr	
	Background Shape &    ${\pm}0.5$\cr	
	Jet and Lepton Efficiencies &     ${}^{+0.1}_{-0.2}$\cr
	Pileup &  $<$0.1 \cr
	QCD effects &     $\pm$0.6\cr\hline
	Total & ${}^{+1.7}_{-2.1}$\cr\hline
	\end{tabular}
	}
	\end{center}
	\label{t:systematics}
\end{table}

\section{Results and Discussion}
\label{s:results}
\begin{figure*}[htbp]
	\begin{center}
		\includegraphics[width=\textwidth]{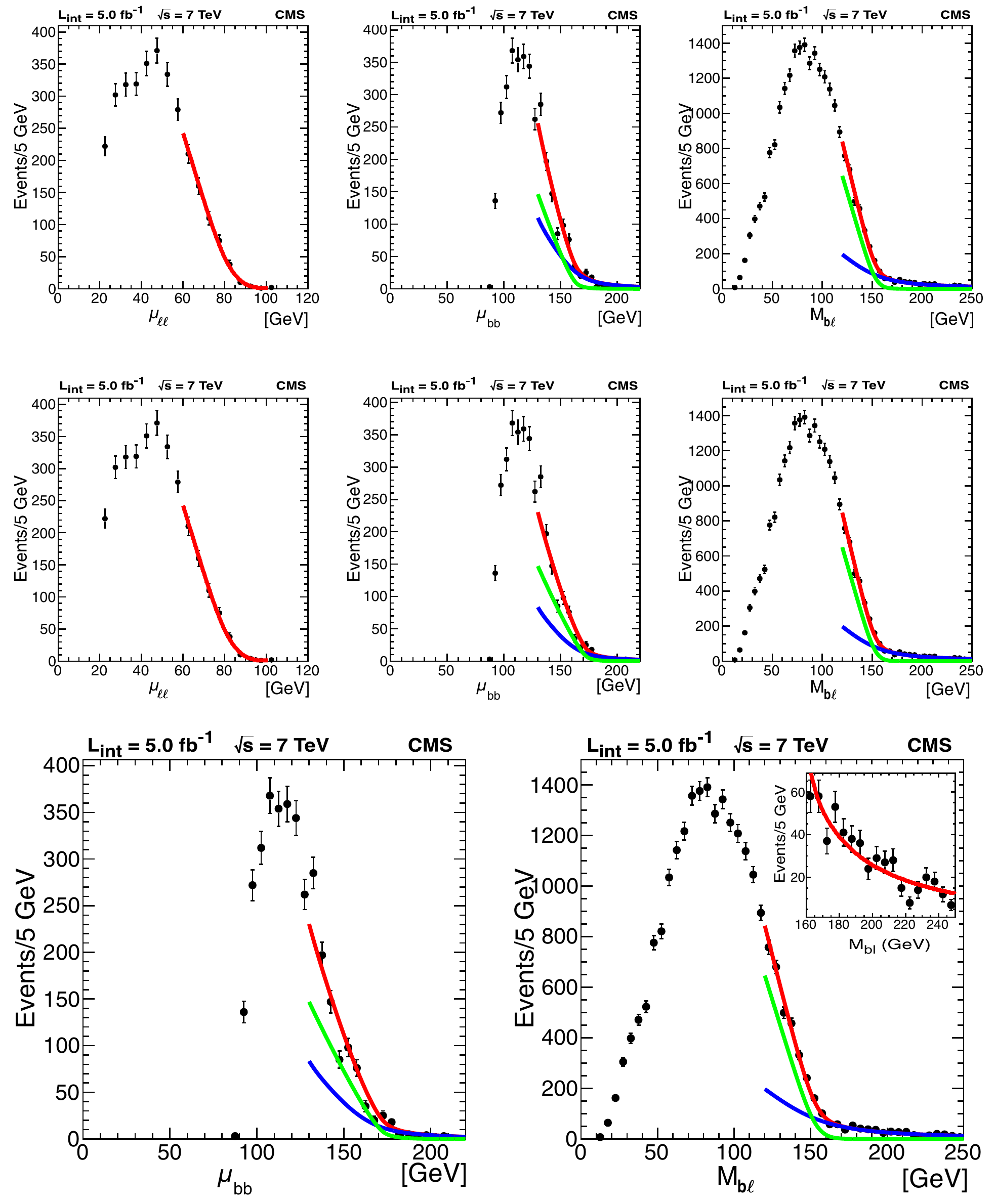}
		\caption{
		Results of simultaneous fits to $\mn^2$, \mw, and \mt. The upper red line
		is in all cases the full fit, while the green (middle) and blue (lowest)
		curves are for the signal and background shapes, respectively. While the
		fit is performed event-by-event for all measured kinematic values, the line
		shown is an approximate extrapolation of the total fit likelihood function
		over the entire fit range. Top row: unconstrained fit; Middle row:
		singly-constrained fit; Bottom row: doubly-constrained fit. The inset shows
		a zoom of the tail region in \mbl\ for the doubly-constrained case to
		illustrate the level of agreement between the background shape and the data
		points.
		}
		\label{f:simfit}
	\end{center}
\end{figure*}

The simultaneous fit to the three distributions  determines \mnsq, \mw, and
\mt. A complete summary of central values and statistical and systematic
uncertainties for all three mass constraints can be found in
Table~\ref{t:results}. Figure~\ref{f:simfit} shows the corresponding fits.

\begin{table*}[htbp]
	\topcaption{
	Fit results from the three mass analyses with various mass constraints.
	Uncertainties are statistical (first) and systematic (second). Values
	in parentheses are constrained in the fit. For the neutrino, squared
	mass is the natural fit variable -- see text for discussion.}
	\begin{center}
	{\renewcommand{\arraystretch}{1.4}
	\begin{tabular}{cccc}
	\hline&\multicolumn{3}{c}{Constraint}\cr
	\cline{2-4}
	\multicolumn{1}{c}{Fit quantity}    & \multicolumn{1}{c}{None}& \multicolumn{1}{c}{$\mn=0$} & \multicolumn{1}{c}{$\mn=0$ and $\mw=80.4\GeV$}                          \cr\hline	
	\mnsq\ ($\GeVns^{\, 2}$) & $-556\pm 473 \pm 622$ & (0)                             & (0)                          \cr
	\mw\ (\GeVns{})         & $72  \pm 7   \pm 9$   & $80.7 \pm 1.1 \pm 0.6$          & (80.4)                          \cr
	\mt\ (\GeVns{})         & $163 \pm 10  \pm 11$  & $174.0\pm 0.9 {}\oursyst $ & $173.9\pm 0.9 {}\oursyst $\cr\hline
	\end{tabular}
	}
	\end{center}
	\label{t:results}
\end{table*}

We take the doubly-constrained version to be the final result:
\begin{equation}
	\label{e:raw}
	\mt = 173.9 \pm 0.9\stat{}\oursyst\syst\GeV.
\end{equation}

In the more general case of the unconstrained measurement, the performance
of the endpoint method illustrated here in the \ttbar\ dilepton system
suggests the technique will be a viable option for mass measurements in a
variety of new-physics scenarios. The precision on \mt\ given by the
doubly-constrained fit, for example, is indicative of the precision with
which we might determine the masses of new colored particles (like
squarks), \textit{as a function} of the input test mass \testmassnu.  Of
course, as shown in the second column of Table~\ref{t:results}, the input
mass \mn\ itself will be determined less precisely. Another plausible
scenario is one in which new physics mimics the leptonic decay of the W
boson.  This can arise in SUSY with $R$-parity violation and a lepton-number
violating term in the superpotential. In this case, the lightest
superpartner could be the charged slepton, which decays to a lepton and
neutrino, just like the SM W boson. Current bounds from LEP indicate that
the slepton must be heavier than 100\GeV.  Given the ${\sim}1$\GeV\
precision provided by the singly-constrained fit on the W boson mass, the W
boson can easily be discriminated from such an object based on its mass.

It is interesting to note also that in the unconstrained case, one can
restrict the range of the neutrino mass (which is treated as an unknown
parameter) reasonably well, within approximately 20\GeV, in line with
previous expectations~\cite{Chang:2009dh}. If the \MET signal is due to SM
neutrinos, rather than heavy WIMPs with masses of order 100\GeV, this
level of precision is sufficient to distinguish the two cases.  If, on the
other hand, the \MET signal is indeed due to heavy WIMPs, one might expect
that the precision on the WIMP mass determination will be no worse than
what is shown here for the neutrino, assuming comparable levels of signal
and background.

\section{Conclusions}
A new technique of mass extraction has been applied to \ttbar\ dilepton
events. Motivated primarily by future application to new-physics scenarios,
the technique is based on endpoint measurements of new kinematic variables.
The three mass parameters \mnsq, \mw, and \mt\ are obtained in a
simultaneous fit to three endpoints. In an unconstrained fit to the three
masses, the measurement confirms the utility of the techniques proposed for
new-physics mass measurements. When \mnsq\ and \mw\ are constrained to $0$
and $80.4\GeV$ respectively, we find $\mt=173.9 \pm 0.9\stat{}\oursyst\syst\GeV$, comparable to other dilepton
measurements. This is the first measurement of the top-quark mass with an
endpoint method. In addition to providing a novel approach to a traditional
problem, it achieves a precision similar to that found in standard methods,
and its use lays a foundation for application of similar methods to future
studies of new physics.

\section*{Acknowledgements}

{\tolerance=500
\hyphenation{Bundes-ministerium Forschungs-gemeinschaft Forschungs-zentren} We congratulate our colleagues in the CERN accelerator departments for the excellent performance of the LHC and thank the technical and administrative staffs at CERN and at other CMS institutes for their contributions to the success of the CMS effort. In addition, we gratefully acknowledge the computing centres and personnel of the Worldwide LHC Computing Grid for delivering so effectively the computing infrastructure essential to our analyses. Finally, we acknowledge the enduring support for the construction and operation of the LHC and the CMS detector provided by the following funding agencies: the Austrian Federal Ministry of Science and Research and the Austrian Science Fund; the Belgian Fonds de la Recherche Scientifique, and Fonds voor Wetenschappelijk Onderzoek; the Brazilian Funding Agencies (CNPq, CAPES, FAPERJ, and FAPESP); the Bulgarian Ministry of Education, Youth and Science; CERN; the Chinese Academy of Sciences, Ministry of Science and Technology, and National Natural Science Foundation of China; the Colombian Funding Agency (COLCIENCIAS); the Croatian Ministry of Science, Education and Sport; the Research Promotion Foundation, Cyprus; the Ministry of Education and Research, Recurrent financing contract SF0690030s09 and European Regional Development Fund, Estonia; the Academy of Finland, Finnish Ministry of Education and Culture, and Helsinki Institute of Physics; the Institut National de Physique Nucl\'eaire et de Physique des Particules~/~CNRS, and Commissariat \`a l'\'Energie Atomique et aux \'Energies Alternatives~/~CEA, France; the Bundesministerium f\"ur Bildung und Forschung, Deutsche Forschungsgemeinschaft, and Helmholtz-Gemeinschaft Deutscher Forschungszentren, Germany; the General Secretariat for Research and Technology, Greece; the National Scientific Research Foundation, and National Office for Research and Technology, Hungary; the Department of Atomic Energy and the Department of Science and Technology, India; the Institute for Studies in Theoretical Physics and Mathematics, Iran; the Science Foundation, Ireland; the Istituto Nazionale di Fisica Nucleare, Italy; the Korean Ministry of Education, Science and Technology and the World Class University program of NRF, Republic of Korea; the Lithuanian Academy of Sciences; the Mexican Funding Agencies (CINVESTAV, CONACYT, SEP, and UASLP-FAI); the Ministry of Science and Innovation, New Zealand; the Pakistan Atomic Energy Commission; the Ministry of Science and Higher Education and the National Science Centre, Poland; the Funda\c{c}\~ao para a Ci\^encia e a Tecnologia, Portugal; JINR (Armenia, Belarus, Georgia, Ukraine, Uzbekistan); the Ministry of Education and Science of the Russian Federation, the Federal Agency of Atomic Energy of the Russian Federation, Russian Academy of Sciences, and the Russian Foundation for Basic Research; the Ministry of Science and Technological Development of Serbia; the Secretar\'{\i}a de Estado de Investigaci\'on, Desarrollo e Innovaci\'on and Programa Consolider-Ingenio 2010, Spain; the Swiss Funding Agencies (ETH Board, ETH Zurich, PSI, SNF, UniZH, Canton Zurich, and SER); the National Science Council, Taipei; the Thailand Center of Excellence in Physics, the Institute for the Promotion of Teaching Science and Technology of Thailand and the National Science and Technology Development Agency of Thailand; the Scientific and Technical Research Council of Turkey, and Turkish Atomic Energy Authority; the Science and Technology Facilities Council, UK; the US Department of Energy, and the US National Science Foundation.

Individuals have received support from the Marie-Curie programme and the European Research Council and EPLANET (European Union); the Leventis Foundation; the A. P. Sloan Foundation; the Alexander von Humboldt Foundation; the Belgian Federal Science Policy Office; the Fonds pour la Formation \`a la Recherche dans l'Industrie et dans l'Agriculture (FRIA-Belgium); the Agentschap voor Innovatie door Wetenschap en Technologie (IWT-Belgium); the Ministry of Education, Youth and Sports (MEYS) of Czech Republic; the Council of Science and Industrial Research, India; the Compagnia di San Paolo (Torino); and the HOMING PLUS programme of Foundation for Polish Science, cofinanced from European Union, Regional Development Fund.
\par}

\bibliography{auto_generated} 

\appendix
\section{Analytical resolution functions for \texorpdfstring{\mttpa}{m[ll]} }

We present the analytical forms of the resolution functions used
in the \mttpa\ fits, together with a brief summary of their derivation.

The leptons used in computing \mttpa\ are approximately massless and
therefore the $M_{T2\perp}$ variable may be written~\cite{Konar:2009wn} as
\ifthenelse{\boolean{cms@external}}{
\begin{equation}\begin{split}
	{\mttpa^2}=&
	2(E_{\text{T1}\perp}
	  E_{\text{T2}\perp}
	-\vec{p}_{\text{T1}\perp}
	\cdot
	\vec{p}_{\text{T2}\perp})
	\\
	=&{2p_{\text{T1}}p_{\text{T2}}}
	\Bigl[
	~
	|\sin(\phi_1-\oPhi)\sin(\phi_2-\oPhi)|-
\\
&
	 \sin(\phi_1-\oPhi)\sin(\phi_2-\oPhi)
	~
	\Bigr],
	\label{eq:A1}
\end{split}
\end{equation}
}{
\begin{equation}\begin{split}
	{\mttpa^2}=&
	2(E_{\text{T1}\perp}
	  E_{\text{T2}\perp}
	-\vec{p}_{\text{T1}\perp}
	\cdot
	\vec{p}_{\text{T2}\perp})
	\\
	=&{2p_{\text{T1}}p_{\text{T2}}}
	\left[
	~
	|\sin(\phi_1-\oPhi)\sin(\phi_2-\oPhi)|-
	 \sin(\phi_1-\oPhi)\sin(\phi_2-\oPhi)
	~
	\right],
	\label{eq:A1}
\end{split}\end{equation}
}
where $(p_{Ti},\phi_i)$ are the transverse coordinates of lepton $i\in\{1,2\}$ and
$\oPhi$ is the azimuthal angle of the upstream momentum in the CMS reference frame.

If the upstream \vPt\ vector happens to lie $between$ the two lepton
vectors $\vpta$ and $\vptb$, so that $\phi_1-\oPhi>0$ and $\phi_2-\oPhi<0$
(or vice-versa) then the value of \mttpa\ is identically zero. This is the
origin of the delta function in Eq.~\ref{e:konstantin}. It is convenient
to measure $\oPhi$ from the midline between the lepton \vpt\ vectors rather
than from the CMS-defined $x$ axis, and hence we define $\nuPhi\equiv
\oPhi-\myhalf(\phi_1+\phi_2)$.  We also define the separation between the
two lepton vectors: $\Dphi\equiv \phi_1-\phi_2$.

Eq. \ref{eq:A1} can now be rewritten as:
\ifthenelse{\boolean{cms@external}}{
\begin{equation}\begin{split}
\mu(\nuPhi)=
	\begin{cases}
		\sqrt{{4p_{\text{T}1}p_{\text{T}2}}\sin(\nuPhi-\hDphi)\sin(\nuPhi+\hDphi)}\\
		\hphantom{0} \qquad\text{when }|\nuPhi|>|\hDphi|;  \\
		0 \qquad\text{otherwise.}
	\end{cases}
	\label{eq:A2}
\end{split}
\end{equation}
}{
\begin{equation}
\mu(\nuPhi)=
	\begin{cases}
		~\sqrt{{4p_{\text{T}1}p_{\text{T}2}}\sin(\nuPhi-\hDphi)\sin(\nuPhi+\hDphi)}
		& ~~\text{when     }|\nuPhi|>|\hDphi|;  \\
		~0 & ~~\text{otherwise.}
	\end{cases}
	\label{eq:A2}
\end{equation}
}
To streamline the notation, we have dropped the subscript $\ell\ell$.
(In any case these remarks apply only to calculations on the $\ell\ell$ system.)

The leptonic observables $p_{\text{T}1}$, $p_{\text{T}2}$, $\phi_1$, and
$\phi_2$ are well-measured compared to the direction of the upstream jets,
$\Phi$, and thus the resolution of $\mu(\nuPhi)$  in a given event depends
only on the $\Phi$ resolution, with the leptonic observables treated as
fixed parameters. The distribution of $\nuPhi$ is well-approximated by a
Gaussian form,  with $\sigma_{\nuPhi}\ll\pi$; we ignore small non-Gaussian
tails.

The functional form given in Eq. \ref{eq:A2} is maximal at
$\nuPhi=\pi/2$, falls to zero on either side at $\nuPhi={\pm}\hDphi$, and is
exactly zero in the neighboring regions $[0,\hpi-\hDphi]$ and
$[\hpi+\hDphi,\pi]$. The function is periodic in $\nuPhi$ with period
$\pi$, but because of the condition $\sigma_{\nuPhi}\ll\pi$ we restrict our
attention to the interval $0\le \nuPhi \le \pi$. For the non-zero portion
of $\mu(\nuPhi)$ there is also an inverse function:
\begin{equation}
	\nuPhi(\mu)=\frac{1}{2} \cos^{-1}
	\left[
	\frac{\mu^{2}_{\max}-\mu^{2}}{2\ptpt} -1
	\right].
	\label{eq:A3}
\end{equation}
The inverse function $\nuPhi(\mu)$ is double-valued as one value of $\mu$
maps to two values of $\nuPhi$ located symmetrically on either side of
$\pi/2$. The maximum value of $\mu$, here denoted $\mu_{\max}$, is the
largest value $\mu$ can take on for the given the lepton momentum vectors;
it corresponds to $\nuPhi=\pi/2$ where the lepton bisector is orthogonal to
the upstream momentum. It should not be confused with the endpoint of the
$\mu$ distribution, which, in addition to the upstream momentum orientation
$\nuPhi=\pi/2$, also requires extreme lepton kinematics:
$p_{\text{T}1}p_{\text{T}2}$  maximal and $\Dphi=0$ (leptons collinear).

To map the Gaussian PDF $G(\nuPhi|\sigma_{\nuPhi})$ into a resolution
function ${R}_{1}(\mu)$, we write:
\begin{equation}
	{R}_{1}(\mu) = \sum {G(\nuPhi(\mu)|\sigma_{\nuPhi})}\left|\frac{\rd\nuPhi(\mu)}{\rd\mu}\right|,
	\label{eq:A4}
\end{equation}
where the sum is over the two branches of the double-valued $\nuPhi(\mu)$.
The derivatives of $\nuPhi(\mu)$ and $\mu(\nuPhi)$ have simple analytic
forms.

In the region where $\mu(\nuPhi)=0$, $R(\mu)$ is a delta function
$R_{0}\delta(\mu)$ whose amplitude $R_{0}$ is given by the area under
$G(\nuPhi|\sigma_{\nuPhi})$ in the angular region between the two leptons,
$R_{0}\equiv\int_{-\Dphi/2}^{\Dphi/2}G(\nuPhi|\sigma_{\nuPhi})\,\rd\nuPhi$.
Thus the total resolution function is given by
\begin{equation}
	{R}(\mu) = R_{0}\delta(\mu) +
	\Theta(\mu)
	\sum {G(\nuPhi(\mu)|\sigma_{\nuPhi})}\left|\frac{\rd\nuPhi(\mu)}{\rd\mu}\right|,
	\label{eq:A5}
\end{equation}
where $\Theta(\mu)$ is the unit step function transitioning from 0 to 1 at $\mu=0$.

\begin{figure}
	\begin{center}
		\includegraphics[width=0.48\textwidth]{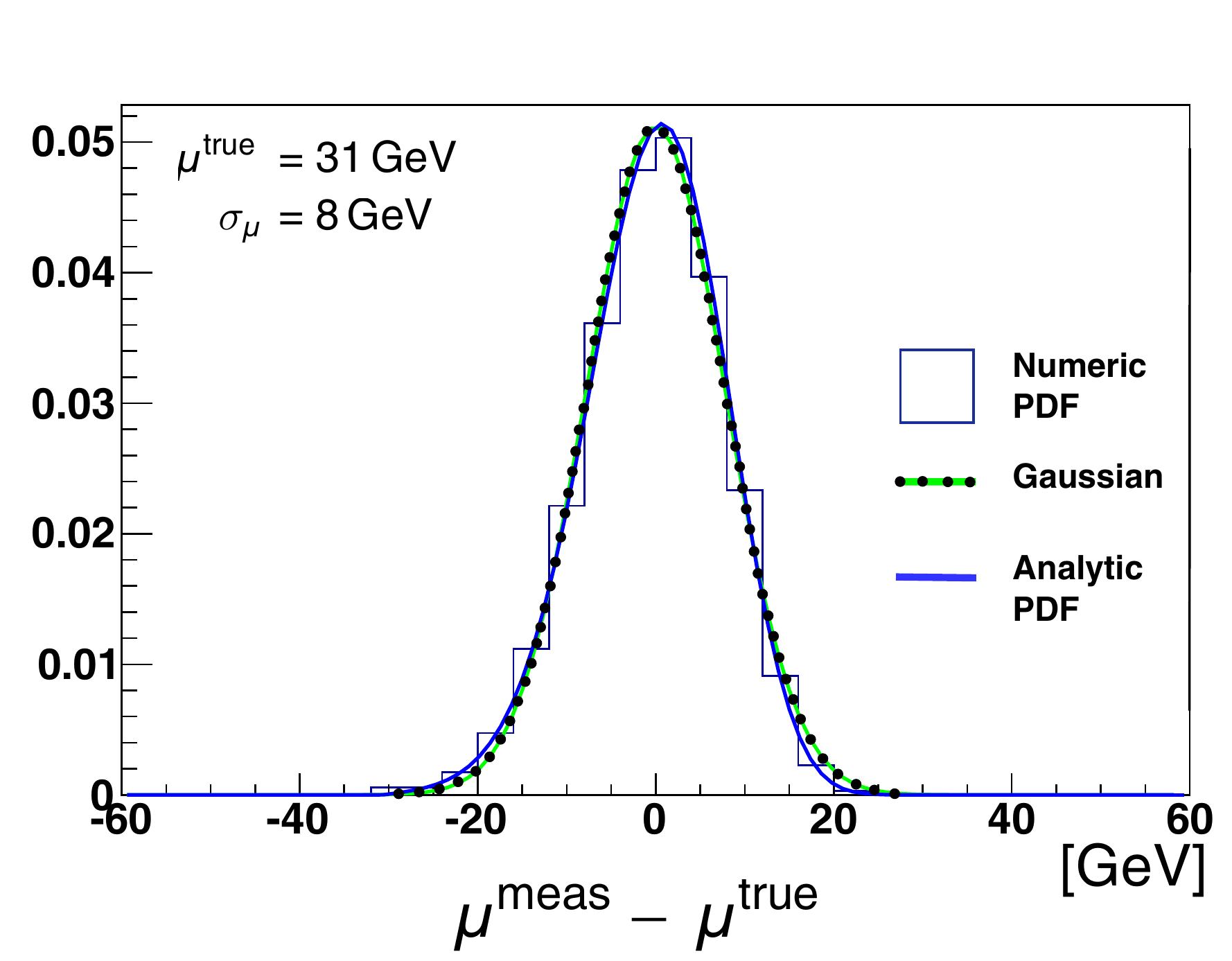}
		\includegraphics[width=0.48\textwidth]{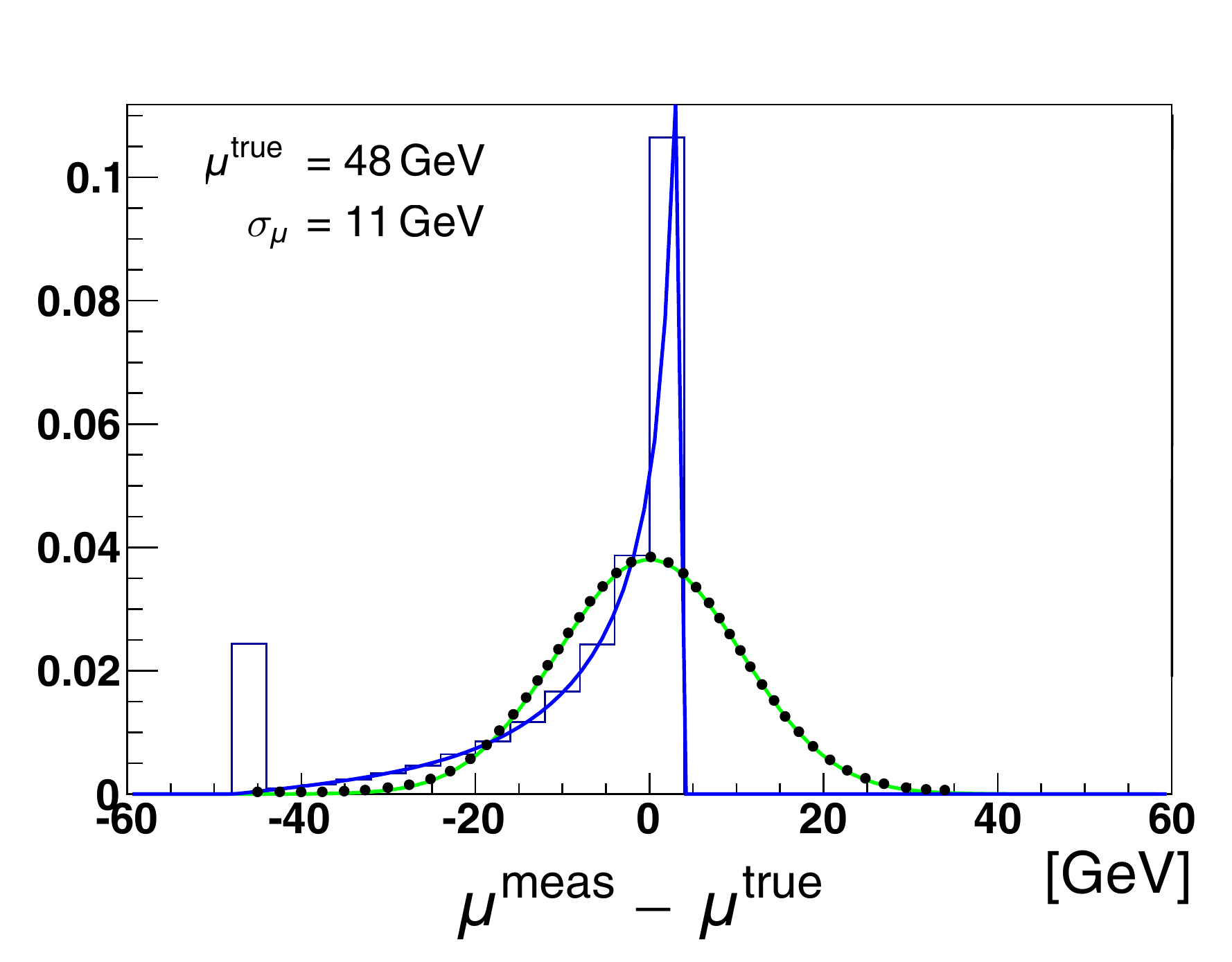}
		\caption{Example resolution functions. The panels show two events with
		different lepton and upstream momentum kinematics, as discussed in the
		text. The dotted curve is a Gaussian with a $\sigma$ given by the linearly
		propagated uncertainties; and the solid curve is the analytic form of the
		resolution function, given in Eq.~(\ref{eq:A4}). The histogram  is created
		by numerically propagating resolutions in the underlying parameters.\label{fig:resfuncs}}
		\end{center}

\end{figure}

Figure~\ref{fig:resfuncs}
shows two representative cases, showing the range of resolution function
behavior from Gaussian to sharply peaked.  In the latter case the delta
function $R_{0}\delta(\mu)$ is not plotted.  In the \cmsLeft panel the $\nuPhi$
is midway between the extremes ${\pm}\hDphi$ and $\pi/2$ and the
$\sigma_{\nuPhi}$ is relatively narrow; in the \cmsRight panel, $\nuPhi$ is
closer to $\pi/2$ and has a large value of $\sigma_{\nuPhi}$ that allows
smearing into the $-\hDphi<\nuPhi<\hDphi$ region.  The high bin at
$-45$\GeV\ in the histogram component of the \cmsRight panel contains events
in which the resolution smearing of the upstream momentum vector pushed the
\mttpa\ value into the delta function at $\mttpa=0$.  In the analytic form,
the corresponding feature would be the delta function $R_0 \delta(\mu)$;
but, as noted above, this has not been explicitly drawn.

\cleardoublepage \section{The CMS Collaboration \label{app:collab}}\begin{sloppypar}\hyphenpenalty=5000\widowpenalty=500\clubpenalty=5000\textbf{Yerevan Physics Institute,  Yerevan,  Armenia}\\*[0pt]
S.~Chatrchyan, V.~Khachatryan, A.M.~Sirunyan, A.~Tumasyan
\vskip\cmsinstskip
\textbf{Institut f\"{u}r Hochenergiephysik der OeAW,  Wien,  Austria}\\*[0pt]
W.~Adam, T.~Bergauer, M.~Dragicevic, J.~Er\"{o}, C.~Fabjan\cmsAuthorMark{1}, M.~Friedl, R.~Fr\"{u}hwirth\cmsAuthorMark{1}, V.M.~Ghete, N.~H\"{o}rmann, J.~Hrubec, M.~Jeitler\cmsAuthorMark{1}, W.~Kiesenhofer, V.~Kn\"{u}nz, M.~Krammer\cmsAuthorMark{1}, I.~Kr\"{a}tschmer, D.~Liko, I.~Mikulec, D.~Rabady\cmsAuthorMark{2}, B.~Rahbaran, C.~Rohringer, H.~Rohringer, R.~Sch\"{o}fbeck, J.~Strauss, A.~Taurok, W.~Treberer-Treberspurg, W.~Waltenberger, C.-E.~Wulz\cmsAuthorMark{1}
% ...
\vskip\cmsinstskip
% ...
1:~~Also at Vienna University of Technology, Vienna, Austria\\
2:~~Also at CERN, European Organization for Nuclear Research, Geneva, Switzerland\\
% ...

\end{sloppypar}
\end{document}